\documentclass[10pt,letterpaper]{amsart}
\pdfoutput=1
\usepackage[utf8]{inputenc}
\usepackage{textcomp,marvosym}
\usepackage{fixltx2e}
\usepackage{amsmath,amssymb}
\usepackage{cite}
\usepackage{nameref}
\makeatletter
\renewcommand{\@biblabel}[1]{\quad#1.}
\makeatother
\newcommand{\fluxunit}{\textmu mol\,m\textsuperscript{-2}\,s\textsuperscript{-1}}

\usepackage{graphicx}
\usepackage{url}
\usepackage[export]{adjustbox}

\usepackage{newfloat}
\DeclareFloatingEnvironment[name={Supporting Figure}]{suppfigure}
\DeclareFloatingEnvironment[name={Supporting Table}]{supptable}

\usepackage{caption}
\makeatletter
\newcommand{\addresseshere}{
\enddoc@text\let\enddoc@text\relax
}
\makeatother

\title[Multiscale metabolic modeling of C4 plants]{Multiscale metabolic modeling of C4 plants: connecting
  nonlinear genome-scale models to leaf-scale metabolism in developing
  maize leaves}
\author[E. Bogart]{Eli Bogart}
\address{\hfill\break Laboratory of Atomic and Solid State Physics/Institute of Biotechnology\hfill\break 
Cornell University\hfill\break
Ithaca, NY\hfill\break
{\tt c.myers@cornell.edu}}
\author[C.R. Myers]{Christopher R. Myers}

\begin{document}
\begin{abstract}
C4 plants, such as maize, concentrate carbon dioxide in a specialized
compartment surrounding the veins of their leaves to improve the
efficiency of carbon dioxide assimilation. Nonlinear relationships
between carbon dioxide and oxygen levels and reaction rates are key to
their physiology but cannot be handled with standard techniques of
constraint-based metabolic modeling.  We demonstrate that
incorporating these relationships as constraints on reaction rates and
solving the resulting nonlinear optimization problem yields realistic
predictions of the response of C4 systems to environmental and
biochemical perturbations.  Using a new genome-scale reconstruction of
maize metabolism, we build an 18000-reaction, nonlinearly constrained
model describing mesophyll and bundle sheath cells in 15 segments of
the developing maize leaf, interacting via metabolite exchange, and
use RNA-seq and enzyme activity measurements to predict spatial
variation in metabolic state by a novel method that optimizes
correlation between fluxes and expression data. Though such
correlations are known to be weak in general, here the predicted fluxes
achieve high correlation with the data, successfully capture the
experimentally observed base-to-tip transition between
carbon-importing tissue and carbon-exporting tissue, and include a
nonzero growth rate, in contrast to prior results from similar methods
in other systems. We suggest that developmental gradients may be
particularly suited to the inference of metabolic fluxes from
expression data.
\end{abstract}

\maketitle
\thispagestyle{empty}


\section*{Introduction}
C4 photosynthesis is an anatomical and biochemical system which
improves the efficiency of carbon dioxide assimilation in plant leaves
by restricting the carbon-fixing enzyme Rubisco to specialized bundle
sheath compartments surrounding the veins, where a
high-CO\textsubscript{2} environment is maintained that favors
CO\textsubscript{2} over $\text{O}_2$ in their competition for Rubisco
active sites, thus suppressing photorespiration
\cite{vonCaemmerer2003}.  

C4 plants are geographically and
phylogenetically diverse, and represent the descendants of over 60
independent evolutionary origins of the system \cite{Sage2011}.  They
include major crop plants such as maize, sugarcane and sorghum as well
as many weeds and, relative to non-C4 (C3) plants, typically show
improved nitrogen and water use efficiencies \cite{Brown1999}.  The
agricultural and ecological significance of the C4 system and its
remarkable convergent evolution have made it the object of intense
study.  The core biochemical pathways are now generally understood
\cite{Kanai1999} but many areas of active research remain, including
the genetic regulation of the C4 system \cite{Hibberd2010}, the
importance of particular components of the system to its function
(e.g., \cite{Studer2014}), the significance of inter-specific
variations in C4 biochemistry \cite{Furbank2011}, details of the
process of C4 evolution, \cite{Sage2004, Christin2009, Griffiths2013,
  Heckmann2013, Way2014} and the prospect of increasing yields of C3
crops by artificially introducing C4 functionality to those species
\cite{Covshoff2012, vonCaemmerer2012}.

Computational and mathematical modeling is a proven approach to
gaining insight into C4 photosynthesis and will play an important role
in addressing these questions.  High-level nonlinear models of
photosynthetic physiology \cite{VonCaemmerer2000} relating enzyme
activities, light and atmospheric CO\textsubscript{2} levels, and the
rates of CO\textsubscript{2} assimilation by leaves have been widely
applied to infer biochemical properties from macroscopic experiments
and explore the responses of C4 plants under varying conditions.  More
recently, detailed kinetic models have been used to explore the
optimal allocation of resources to enzymes in an NADP-ME type C4 plant
\cite{Wang2014} and the relationship between the three decarboxylation
types \cite{Wang2014a}.

Large-scale constraint-based metabolic models offer particular
advantages for the investigation of connections between the C4 system
and a plant's metabolism more broadly (for example, partitioning of
nonphotosynthetic functions between mesophyll and bundle sheath, or
the evolutionary recruitment of nonphotosynthetic reactions into the
C4 cycle) and for interpreting high-throughput experimental data from C4
systems. 
Photosynthesis is difficult to describe, however, using the standard
approach of flux balance analysis (FBA), which predicts reaction rates $v_1,
v_2, \ldots v_N$ in a metabolic network by optimizing a biologically
relevant function of the rates subject to the requirement that the
system reach an internal steady state,
\begin{equation}
  \label{FBA}
  \begin{aligned}
    & \max_{(v_1, v_2, \ldots, v_N) \in \mathbb R^N} & & f(\mathbf v) \\
    & \text{s.t.} & & S\cdot\mathbf v = \mathbf 0,\\
  \end{aligned}
\end{equation}
where the stoichiometry matrix $S$ is determined by the network
structure \cite{Orth2010}.  The relationship between the rate $v_c$
of carbon fixation by Rubisco and the rate $v_o$ of the Rubisco
oxygenase reaction depends nonlinearly on the ratio of the local
oxygen and carbon dioxide concentrations (here expressed as equivalent
partial pressures),
\begin{equation}\label{vo_vc_ratio}
  \frac{v_o}{v_c} = \frac {1}{S_R} \frac{P_{O2}}{P_{CO2}}
\end{equation}
where $S_R$ is the specificity of Rubisco for CO\textsubscript{2} over
O\textsubscript{2}.  In the C4 case, the CO\textsubscript{2} level in
the bundle sheath compartment is itself a function of the rates of the
reactions of the C4 carbon concentration system and the rate of
diffusion of CO\textsubscript{2} back to the mesophyll.

With the addition of (\ref{vo_vc_ratio}), the problem (\ref{FBA})
becomes nonlinear and cannot be solved with typical FBA tools; instead
(as the problem is also nonconvex \cite{Boyd2004}), a general-purpose
nonlinear programming algorithm is required to numerically solve it.

Prior constraint-based models of plant metabolism have typically
ignored the constraint (\ref{vo_vc_ratio}) or assumed the oxygen and
carbon dioxide levels $P_{O2}$ and $P_{CO2}$ are known and fixed
$v_o/v_c$ accordingly \cite{deOliveiraDalMolin2010,Saha2011}.  While
this approach is suitable for mature C4 leaves under many conditions,
where $v_o/v_c$ is approximately zero, it may break down in some of
the most important targets for simulation: developing tissue, mutants,
and C3-C4 intermediate species, where $P_{CO2}$ in the bundle sheath
compartment is not necessarily high.

In other recent work, a high-level physiological model was used to
determine $v_o$, $v_c$, and other key reaction rates given a few
parameters, which were then fixed in order to solve
eq. (\ref{FBA})\cite{Heckmann2013} .  This method yields realistic
solutions, but its application is limited by the lack of a way to set
the necessary phenomenological parameters (e.g., the maximum rate of
PEP regeneration in the C4 cycle) based on lower-level, per-gene data
(e.g., from transcriptomics or experiments on single-gene mutants).

Here, we treat the problem in a more general way by incorporating the
nonlinear constraint (\ref{vo_vc_ratio}) directly into the
optimization problem (\ref{FBA}) and solving the resulting nonlinear
program numerically with the IPOPT package \cite{Wachter2006}, using a
new computational interface that we have developed, which allows
rapid, interactive development of nonlinearly-constrained FBA problems
from metabolic models specified in SBML format \cite{Hucka2003}.

Using a new genome-scale reconstruction of the metabolic network of
\textit{Zea mays}, developed with particular attention to
photosynthesis and related processes, we confirm that this approach
can reproduce the nonlinear responses of well-validated, high-level
physiological models of C4 photosynthesis \cite{VonCaemmerer2000},
while also providing detailed predictions of fluxes throughout the
network.

\begin{figure}
  \includegraphics[width=1.25\textwidth,center]{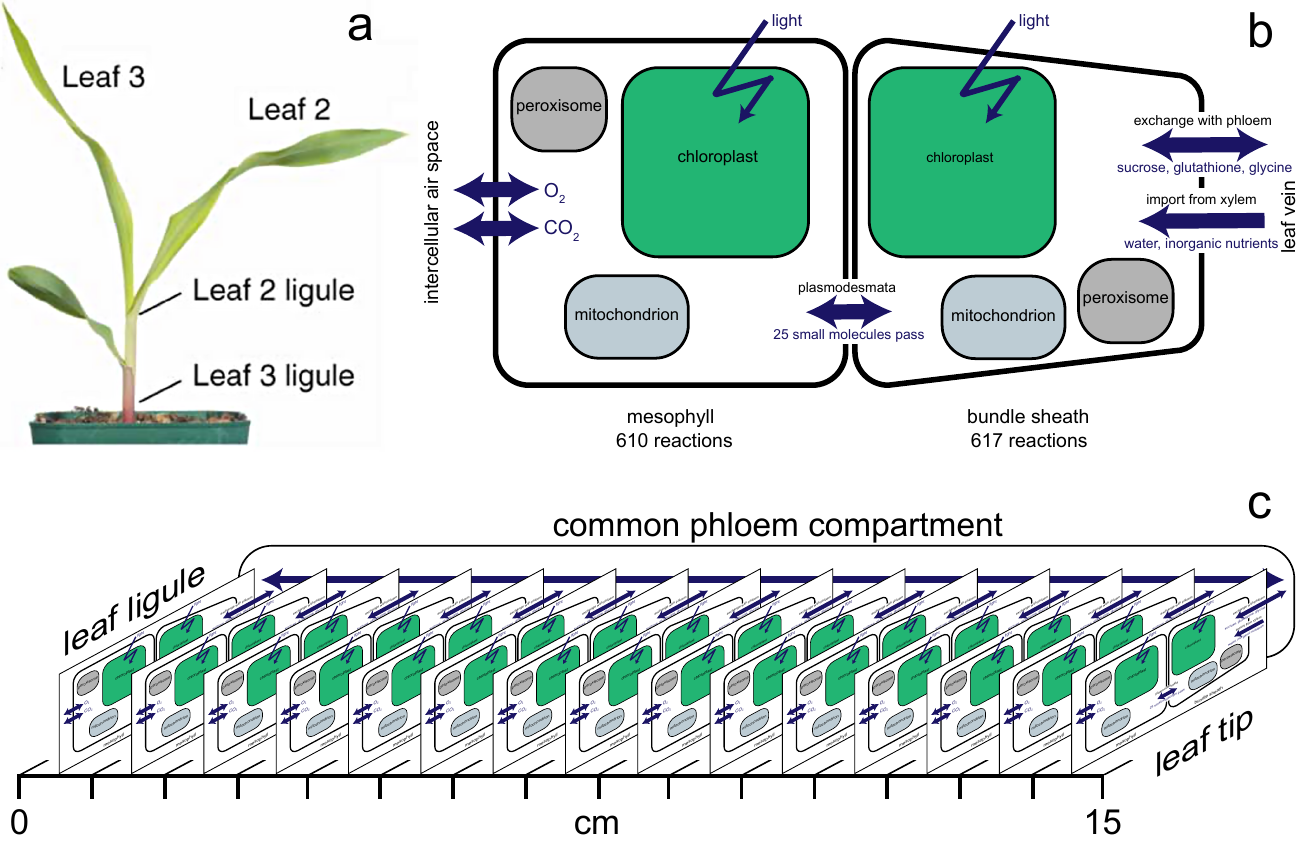}
  \caption{{\bf Maize plant and models.} (a) Nine-day-old maize plant
    (image from \cite{Li2010}).  (b) Organization of the two-cell-type
    metabolic model, showing compartmentalization and exchanges across
    mesophyll and bundle sheath cell boundaries.  (c) Combined
    121-compartment model for leaf 3 at the developmental stage shown
    in (a). Fifteen identical copies of the model shown in (b)
    represent 1-cm segments from base to tip.}
\label{schematicfig}
\end{figure}

Finally, we combine the results of enzyme assay measurements and
multiple RNA-seq experiments and apply a new method to infer the
metabolic state at points along a developing maize leaf
(Fig. \ref{schematicfig}a) using a model of mesophyll and bundle
sheath tissue in fifteen segments of the leaf, interacting through
vascular transport of sucrose, glycine, and glutathione.  We compare
our results to radiolabeling experiments. 

\section*{Results}
\subsection*{Metabolic reconstruction of \textit{Zea mays}}
A novel genome-scale metabolic model was generated from version 4.0 of
the CornCyc metabolic pathway database \cite{CornCyc} and is presented
in two forms. The comprehensive reconstruction involves {2720}
reactions among {2725} chemical species, and incorporates CornCyc
predictions for the function of 5204 maize genes, with 2064 reactions
associated with at least one gene. A high-confidence subset of the
model, excluding many reactions not associated with manually curated
pathways or lacking computationally predicted gene assignments as well
as all reactions which could not achieve nonzero flux in FBA
calculations, involves {635} reactions among {603} species, with 469
reactions associated with a total of 2140 genes.

Both the comprehensive and high-confidence models can simulate the
production of all major maize biomass constituents (including amino
acids, nucleic acids, fatty acids and lipids, cellulose and
hemicellulose, starch, other carbohydrates, and lignins, as well as
chlorophyll) under either heterotrophic or photoautotrophic conditions
and include chloroplast, mitochondrion, and peroxisome compartments,
with key reactions of photosynthesis (including a detailed
representation of the light reactions), photorespiration, the NADP-ME
C4 cycle, and mitochondrial respiration localized appropriately.  Gene
associations for reactions present in more than one subcellular
compartment have been refined based on the results of subcellular
proteomics experiments and computational predictions (as collected by
the Plant Proteomics Database, \cite{Sun2009}) to assign genes to
reactions in appropriate compartments.

A model for interacting mesophyll and bundle sheath tissue in the leaf
was created by combining two copies of the high-confidence model, with
transport reactions to represent oxygen and CO\textsubscript{2} diffusion and
metabolite transport through the plasmodesmata, and restricting
exchange reactions appropriately (nutrient uptake from the vascular
system to the bundle sheath, and gas exchange with the intercellular
airspace to the mesophyll).  A schematic of the two-cell model is
shown in Fig.~\ref{schematicfig}b.

Both single-cell versions of the model and the two-cell model,
designated iEB5204, iEB2140, and iEB2140x2 respectively (based on the
primary author's initials and number of genes included, according to
the established naming convention~\cite{Reed2003}), are available in
SBML format (\nameref{S11_Model}-\nameref{S13_Model}.)

\subsection*{Nonlinear flux-balance analysis}
To solve nonlinear optimization problems incorporating the constraints
discussed above, we developed a Python package which -- given a model in
SBML format, arbitrary nonlinear constraints, a (potentially
nonlinear) objective function, and all needed parameter values -- infers
the conventional FBA constraints of eq.~(\ref{FBA}) from the
structure of the network, automatically generates Python code to
evaluate the objective function, all constraint functions, and their
first and second derivatives, and calls IPOPT through the pyipopt
interface \cite{pyipopt}.  Source code for the package is available in
\nameref{S14_Protocol} and online (\url{http://github.com/ebogart/fluxtools}). The software
has been used to successfully solve nonlinear FBA problems with over
{84000} variables and {62000} constraints.

\begin{figure}
  \includegraphics[width=\textwidth,center]{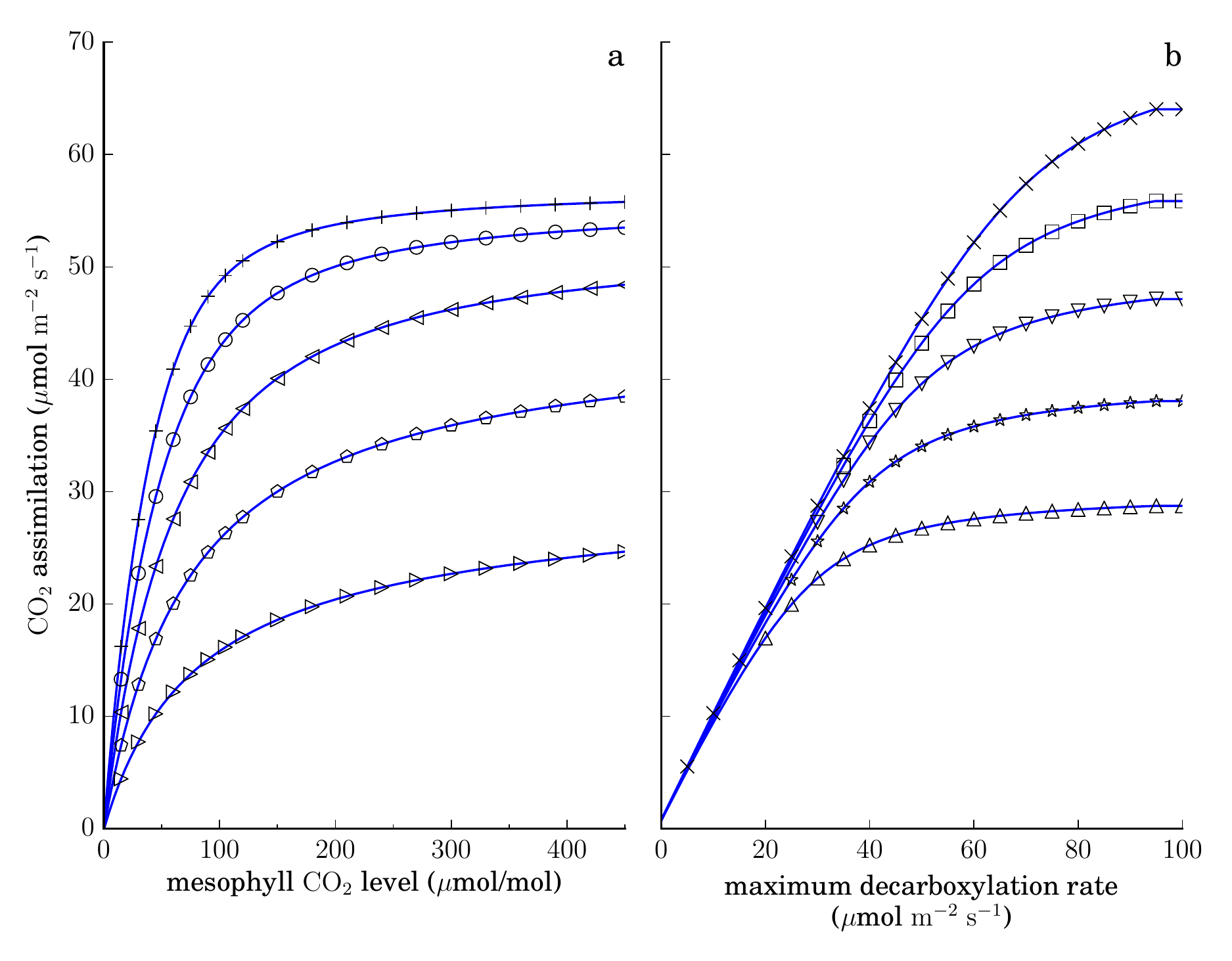}
  \caption{{\bf $\text{CO}_2$ assimilation rates ($A$) predicted by
      the C4 photosynthesis model of \cite{VonCaemmerer2000}, solid
      lines, and the present nonlinear genome-scale model (markers)
      maximizing $\text{CO}_2$ assimilation with equivalent
      parameters.}  Left, $A$ vs mesophyll $\text{CO}_2$ levels with
    varying PEPC levels (top to bottom, $V_{p,\text{max}}=$ 110, 90,
    70, 50, and 30 \fluxunit{}).  Right, $A$ vs
    total maximum activity of all bundle sheath decarboxylase enzymes
    (equivalent to the maximum PEP regeneration rate $V_{pr}$) at
    varying Rubisco levels (top to bottom, $V_{c,\text{max}}=$ 70, 60,
    50, 40, and 30 \fluxunit{}).  Other
    parameters as in Table 4.1 of \cite{VonCaemmerer2000}, except with
    nonphotorespiratory respiration rates $r_d=r_m=0$.}
\label{comparisonfig}
\end{figure}

Figure~\ref{comparisonfig} demonstrates that, as expected, optimizing
the rate of CO\textsubscript{2} assimilation in the two-cell-type
model with nonlinear kinetic constraints
[eqs. (\ref{rubisco_kinetics}), (\ref{pepc_kinetics}),
(\ref{leakiness})] produces predictions consistent with the results of
the physiological model of \cite{VonCaemmerer2000}. Note that the
effective value of one macroscopic physiological parameter may be
governed by many microscopic parameters in the genome-scale
model. In the figure, the effective maximum PEP regeneration rate
$V_{pr}$ is controlled by the maximum rate of three decarboxylase
reactions in the bundle sheath compartment, but with an appropriate
choice of parameter values any of at least 10 reactions of the C4
system could become the rate-limiting step in PEP regeneration, and in
the calculations below, expression levels for any of the 42 genes
associated with these reactions (\autoref{S10_Table}) could influence
the net PEP regeneration rate.

\subsection*{Flux predictions in the developing leaf based on multiple
  data channels}
Maize leaves display a developmental gradient along the base-to-tip
direction, with young cells in the immature base and fully
differentiated cells at the tip~\cite{Li2010, Nelson2011}.  To explore
variations in metabolic state along this axis, we combined the RNA-seq
datasets of Wang et al.~\cite{fifteensegment} and Tausta et
al.~\cite{Tausta2014} to estimate expression levels (as FPKM) for
39634 genes in the mesophyll and bundle sheath cells at 15 points,
representing 1 cm segments of the third leaf of a 9-day-old maize
plant, which includes a full gradient of developmental stages.  The
combined dataset provides expression information for 920 reactions in
the two-cell model (460 each in mesophyll and bundle sheath cells).

A whole-leaf metabolic model, iEB2140x2x15, was created from fifteen
copies of the two-cell model, each representing a 1-cm segment,
interacting through the exchange of sucrose, glycine, and glutathione
through a common compartment representing the phloem. The resulting
121-compartment model, Fig.~\ref{schematicfig}c, involves {18780} reactions among
{16575} metabolites.

Subject to the requirements that reaction rates in each of the 15
segments obey both the FBA steady-state constraints (eq. \ref{FBA})
and the nonlinear constraints governing Rubisco kinetics
(eqs. \ref{rubisco_kinetics}, \ref{leakiness}, and
\ref{pepc_kinetics}, presented in detail below) we determined the set
of rates $v_{ij}$ for each reaction $i$ at each segment $j$ which were
most consistent with the base-to-tip variation in the gene expression
data, by optimizing the objective function
\begin{equation}\label{fitting}
  F(v) = \sum_{i=0}^{N_r}\sum_{j=1}^{15} \frac{\left(e^{s_i}\left|v_{ij}\right|-d_{ij}\right)^2}
                                                                              {\delta^2_{ij}} + 
              \alpha \sum_{i=0}^{N_r} s_i^2
\end{equation}
where $N_r=920$ is the number of reactions associated with at least
one gene present in the expression data, $d_{ij}$ and $\delta_{ij}$
are the expression data and associated experimental uncertainty for
reaction $i$ at leaf segment $j$, and $s_i$ is an optimizable scale
factor associated with reaction $i$.

Effectively, this calculation -- similar to the method of Lee et
al.~\cite{Lee2012} or FALCON~\cite{Barker2014} -- performs a
constrained least-squares fit of the fluxes to the expression data.
Allowing the scale factors $s_i$ to vary emphasizes agreement between
fluxes and data in their trend along the developmental gradient,
rather than in their absolute value: if the data associated with
reaction $R_i$ has average value 100 FPKM, a solution in which $R_i$
has mean flux 10 \fluxunit{} but correlates
well with the data can achieve (with appropriate choice of scale
factor) a lower cost than a solution in which $R_i$ has mean flux 100
\fluxunit{} but is anticorrelated.  The penalty
term $\alpha \sum s_i^2$
favors solutions in which, generally, reactions with larger associated
expression data carry higher fluxes.  The parameter $\alpha$
controlling the tradeoff between these criteria was set arbitrarily to
1.0 in the work presented here.  We require $s_a = s_b$ if reactions
$a$ and $b$ are mesophyll and bundle sheath instances of the same
reaction.

To constrain the overall scale of the fluxes and further improve
accuracy, we incorporated enzyme activity assay data from
\cite{fifteensegment} for seventeen enzymes (including Rubisco and
PEPC) along the 15 leaf segments as additional constraints on the optimization problem, requiring
for each enzyme $k$ and segment $j$
\begin{equation}
  \label{enzymeequation}
  E_{jk} \geq \left| v_{k1}\right| + \ldots + \left| v_{kn}\right|
\end{equation}
where $E_{jk}$ is the measured maximal activity of the enzyme at that
segment and the sum on the right hand side includes all the reactions
which represent enzyme $k$ in the mesophyll, bundle sheath, and
subcompartments of those cells if applicable.

Solving the optimization problem yielded predictions for reaction rates
and other variables (\nameref{S16_Table}).  Upper and lower bounds on
selected variables (\nameref{S17_Table}) were determined through flux
variability analysis (FVA)~\cite{Mahadevan2003}, allowing the
objective function to increase by 0.1\% from its optimal value.


\subsubsection*{Predicted source-sink transition}
\begin{figure}[p]
  \includegraphics{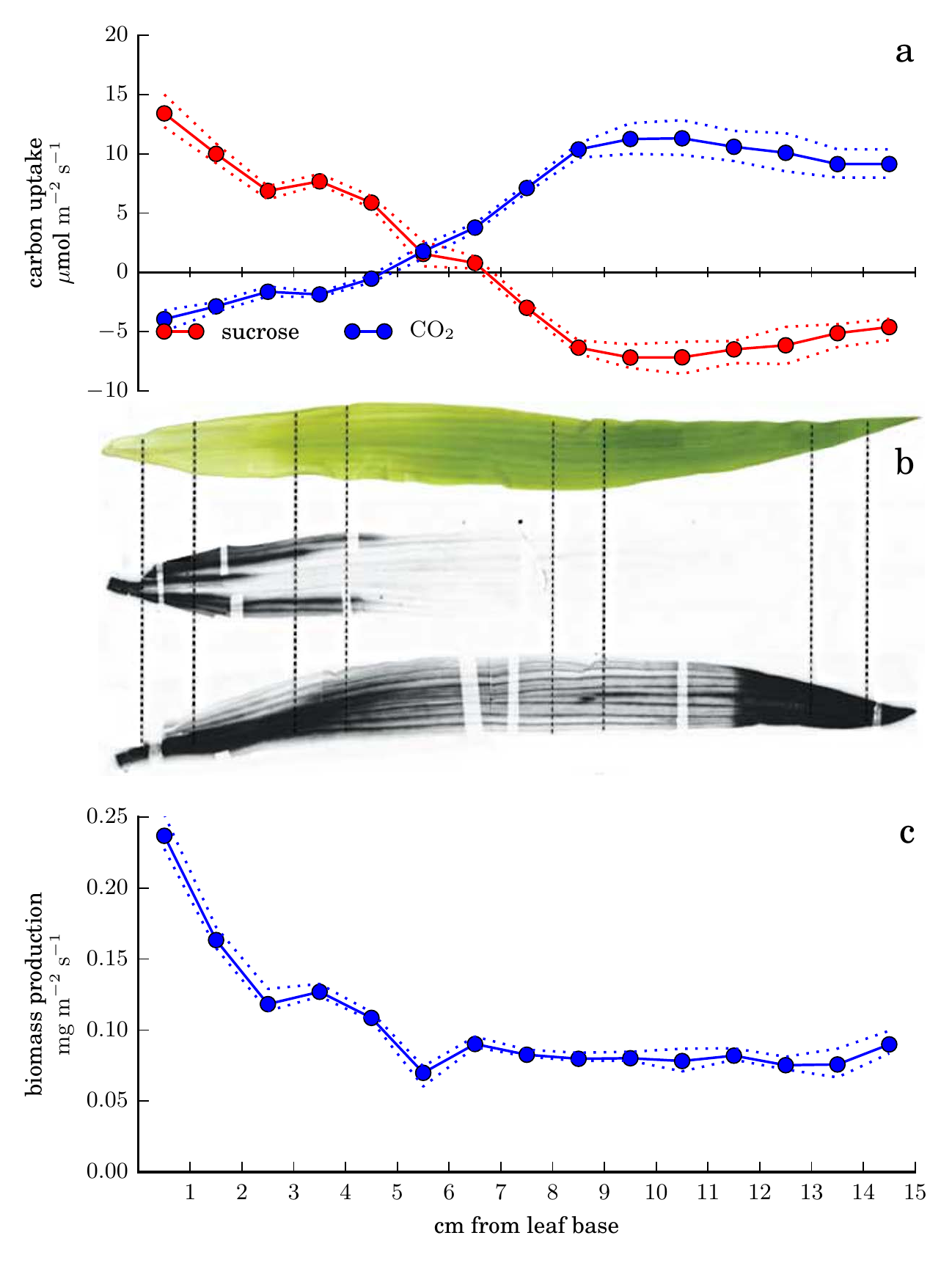}
  \caption{{\bf Source-sink transition along the leaf as predicted by
      optimizing the agreement between fluxes in the nonlinear model
      and RNA-seq data.} (a) Predicted rates of exchange of carbon
    with the atmosphere and phloem along the leaf. (b) Experimental
    observation of the source-sink transition, reproduced
    from~\cite{Li2010}. Upper image, photograph of leaf 3; middle
    image, autoradiograph of leaf 3 after feeding
    \textsuperscript{14}CO\textsubscript{2} to leaf 2; lower image,
    autoradiograph of leaf 3 after feeding
    \textsuperscript{14}CO\textsubscript{2} to the tip of leaf 3.  (c)
    Total biomass production in the best-fitting solution. In panels a
    and c, dotted lines indicate minimum and maximum predicted rates
    consistent with an objective function value no more than 0.1\%
    worse than the optimum.}
\label{sourcesinkfig}
\end{figure}

As shown in Fig.~\ref{sourcesinkfig}, in the outer, more
photosynthetically developed, portion of the leaf, our optimal fit
predicts net CO\textsubscript{2} uptake, with most of the assimilated carbon
incorporated into sucrose and exported to the phloem. Near the base of
the leaf, sucrose is predicted to be imported from the phloem and used
to drive a high rate of biomass production, with some concomitant net
release of CO\textsubscript{2} to the atmosphere by respiration.

This transition between a carbon-exporting source region and a
carbon-importing sink region is well known, and the predicted
transition point between the two, approximately 6 cm above the base of
the leaf, can be compared to the \textsuperscript{14}C-labeling
results of Li et al.~\cite{Li2010} in the same experimental
conditions.  Fig.~\ref{sourcesinkfig}b shows the location of labeled
carbon in leaf 3 after feeding labeled CO\textsubscript{2} to leaf 2
(center image) or leaf 3 (bottom image).  Li et al.~\cite{Li2010}
identified the sink region as the lowest 4 cm of the leaf; the
transition is not perfectly sharp and quantitative comparison of
exchange fluxes is not possible, but the nonlinear FBA results appear
to slightly overestimate the size of the sink region.

Agreement might be improved under a different assumption about net
sucrose import or export by leaf 3 (here, we have assumed that the
import visible in the center image is exactly balanced by the export
suggested by the high density of labeled carbon at the absolute base
in the lower image.)

The net rate of CO\textsubscript{2} assimilation predicted in the
outer, most mature leaf segments, 8-11 \fluxunit{}, is lower than that
typically measured in more mature maize plants (e.g., rates of 20-30
\fluxunit{} in 22-day-old wild-type plants under comparable conditions
\cite{Studer2014}), but photosynthetic capacity may still be
increasing even in these segments.

In addition to sucrose, glycine and glutathione are predicted to be
exported from the source region through the phloem and reimported by
the sink region, consistent with our expectations that nitrogen and
sulfur reduction will occur preferentially in the photosynthesizing
region (\autoref{S1_Figure}).  Note that this behavior emerges from the data
even though there is no explicit requirement in the model that net
phloem transport occur in a basipetal direction.

\subsubsection*{Predicted C4 system function}
\begin{figure}
\includegraphics[width=1.25\textwidth,center]{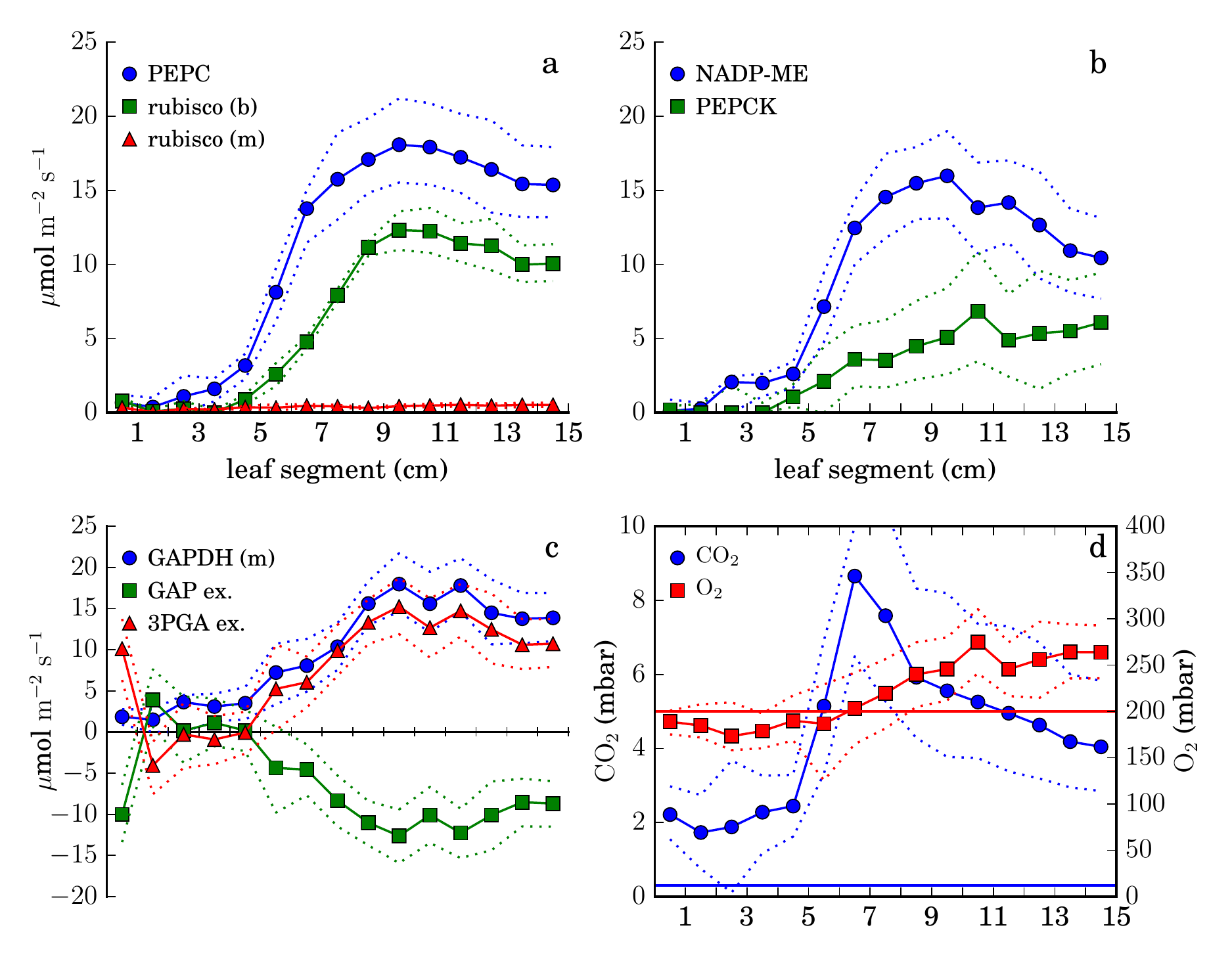}
\caption{{\bf Operation of the C4 system in the best-fitting solution.} (a)
  Rates of carboxylation by PEPC in the mesophyll and Rubisco in the
  mesophyll and bundle sheath.  (b) Rates of $\text{CO}_2$ release by PEP
  carboxykinase and chloroplastic NADP-malic enzyme in the bundle
  sheath. (c) Transport of 3-phosphoglycerate and glyceraldehyde
  3-phosphate from bundle sheath to mesophyll (or the reverse, where
  negative) and glyceraldehyde 3-phosphate dehydrogenation rate in the
  mesophyll chloroplast, showing the involvement of the mesophyll in
  the reductive steps of the Calvin cycle throughout the source
  region. (d) Oxygen and carbon dioxide levels in the bundle
  sheath. Straight lines show mesophyll levels. Throughout, dotted
  lines indicate minimum and maximum predicted values consistent with
  an objective function value no more than 0.1\% worse than the
  optimum.}
\vskip 5.0cm
\label{c4fig}
\end{figure}

Figure \ref{c4fig} shows predicted rates of key reactions of the C4
system and CO\textsubscript{2} and O\textsubscript{2} levels in the bundle sheath.
As expected, the model predicts that a C4 cycle will operate in the
source region of the leaf, elevating the CO\textsubscript{2} level in the
bundle sheath. The CO\textsubscript{2} level is also elevated in the source
region; this is an immediate consequence of respiration in the bundle
sheath and eq.~(\ref{leakiness}).  It may be overestimated here
because we have assumed a constant value for the bundle sheath
CO\textsubscript{2} conductivity (as measured by Bellasio et
al.~\cite{Bellasio2014}); in fact, gene expression associated with
synthesis of the diffusion-resistant suberin layer between bundle
sheath and mesophyll peaks at 4 cm above the leaf base
\cite{fifteensegment}, so $g_s$ is presumably higher below that point.

In the Calvin cycle, most reactions are predicted to be bundle-sheath
specific, but the reductive phase is active in both cells, with
approximately half the 3-phosphoglycerate produced in the bundle
sheath transported to the mesophyll and returned as dihydroxyacetone
phosphate (Fig.~\ref{c4fig}c); this is a known aspect of NADP-ME C4
metabolism connected to reduced photosystem II activity in the bundle
sheath cells \cite{Hatch1987}, which is also predicted here
(\autoref{S2_Figure}).  Consistent with conclusions drawn
independently from the transcriptomic data, as well as proteomic data
from the same system~\cite{Li2010,Majeran2010,fifteensegment}, the
model does not predict a C3-like metabolic state as a developmental
intermediate stage.  As expected in maize \cite{Wingler1999a}, a
significant role for phosphoenolpyruvate carboxykinase (PEPCK) as a
decarboxylating enzyme operating in the bundle sheath in parallel with
NADP-ME is predicted (Fig.~\ref{c4fig}b).

While the predictions are generally consistent with the standard view
of the C4 system in maize, there are minor discrepancies.  In the
mesophyll, our calculations predict that malate production occurs in
the mitochondrion, rather than the chloroplast.  In both mesophyll and
bundle sheath, phosphoenolpyruvate is formed by
pyruvate-orthophosphate dikinase (PPDK) in the chloroplast at a higher
rate than necessary to sustain the C4 cycle; the excess is converted
again to pyruvate by pyruvate kinase in the cytoplasm, with the
resulting ATP consumed by the model's generic ATPase reaction.
Finally, in the bundle sheath, a modest rate of PEPC activity is
predicted, recapturing CO\textsubscript{2} only to have it released
again by the decarboxylases (S3 Figure).  Further refinement of the
associations of genes to reactions in the model might resolve some of
these discrepancies.

\subsubsection*{Global agreement between fluxes and data}
\begin{figure}
\includegraphics[width=1.25\textwidth,center]{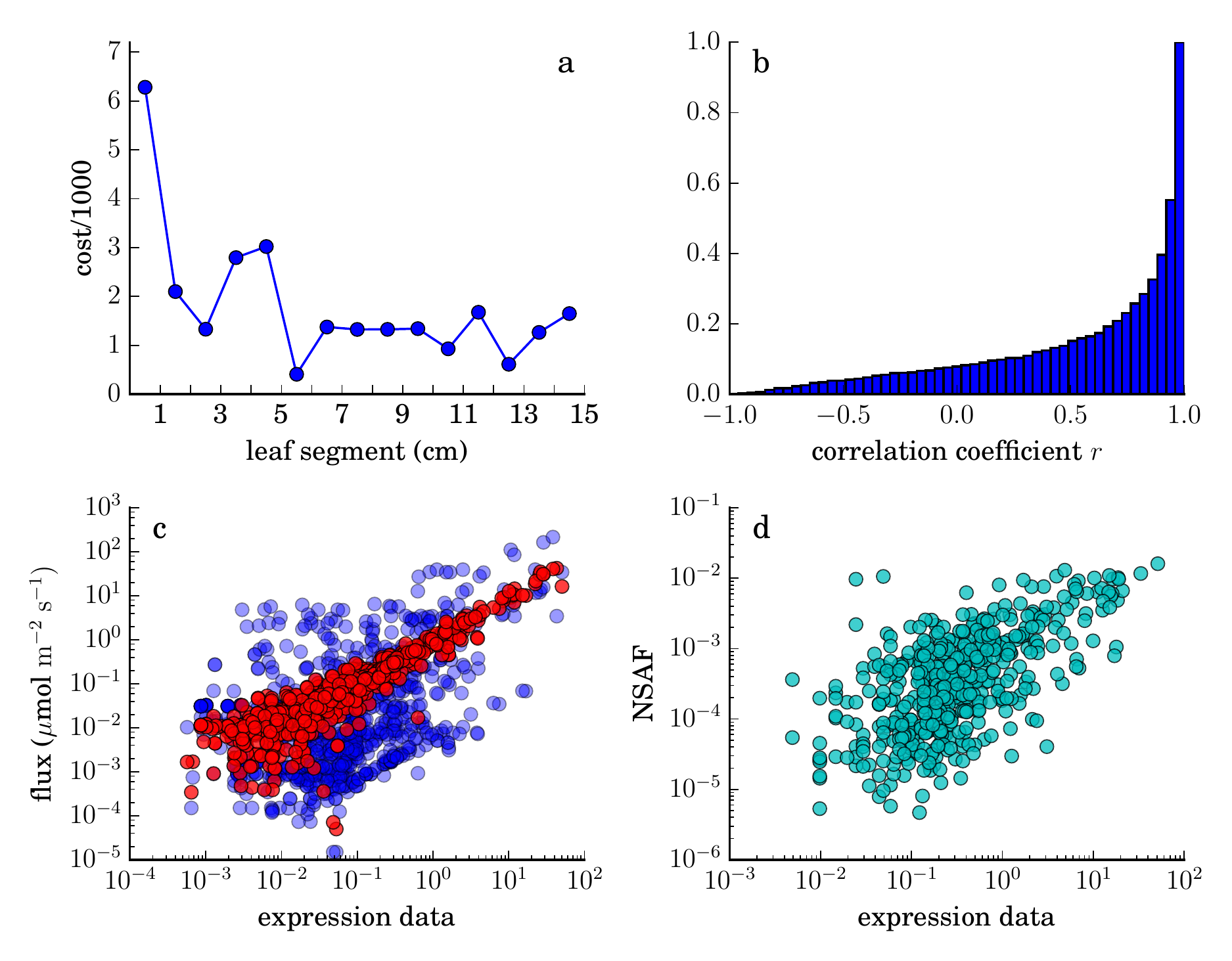}
\caption{{\bf Agreement between RNA-seq data and predicted fluxes.} (a)
  Contribution of each segment to the objective function
  (eq.~(\ref{fitting}), excluding costs associated with scale factors).
  (b) Cumulative histogram of Pearson correlations between data and
  predicted fluxes for all reactions. (c) Predicted fluxes versus
  expression data at the tip of the leaf (blue, raw fluxes; red, after
  rescaling each flux $v_i$ by the optimal factor $e^{s_i}$ of
  eq.~(\ref{fitting})). Some outliers with very low predicted flux are
  not shown. (d) Relationship between RNA-seq and proteomics
  measurements for 506 proteins in the 14th segment from the base,
  redrawn from the data of~\cite{Ponnala2014}. NSAF, normalized
  spectral abundance factor. }
\label{costfig}
\vskip 5.0cm
\end{figure}

Figure \ref{costfig} summarizes overall properties of the predicted
fluxes.  It is not clear why agreement between data and predicted
fluxes is poorer at the base, as shown in Fig.~\ref{costfig}a.  As
discussed below, the cell-type-specific RNA-seq data from Tausta et
al.~\cite{Tausta2014} does not extend below the fourth segment from
the base of the leaf; at the base we have assumed expression levels
for all genes are equal in mesophyll and bundle sheath. Though
proteomics experiments on the same system \cite{Majeran2010} generally
found limited cell-type specificity at the leaf base, this assumption
is likely an oversimplification, and could limit the ability of the
algorithm to find a flux prediction consistent with the data there.

For most reactions, the correlation between the base-to-tip expression
pattern and the base-to-tip trend in predicted flux is high. The
cumulative histogram in Fig.~\ref{costfig}b shows that the Pearson
correlation $r>0.92$ for more than half of the reactions in the model
with associated expression data.

Differences in expression levels between different reactions, however,
correlate only weakly with the differences in fluxes between those
reactions, as shown for segment 15 in Fig.~\ref{costfig}c (blue
circles). After rescaling fluxes by the optimal per-reaction scale
factors, a clear relationship emerges (Fig.~\ref{costfig}c, red
circles), confirming that the scale factors are functioning as
intended.  Of course we should not expect a perfect correlation
between data on transcript levels and predicted fluxes through
associated reactions.  The limited correlation between fluxes and
expression data across different reactions presumably follows, in
part, from the imperfect correlation between expression data and
protein abundance across different genes, as illustrated in
Fig.~\ref{costfig}d with data from the same experimental system
\cite{Ponnala2014}, as well as from the different catalytic
capabilities of different enzymes, posttranslational regulation,
differences in substrate availability, etc.

\subsubsection*{Reconciling expression data and network structure} 
\begin{figure}
  \includegraphics[width=1.25\textwidth,center]{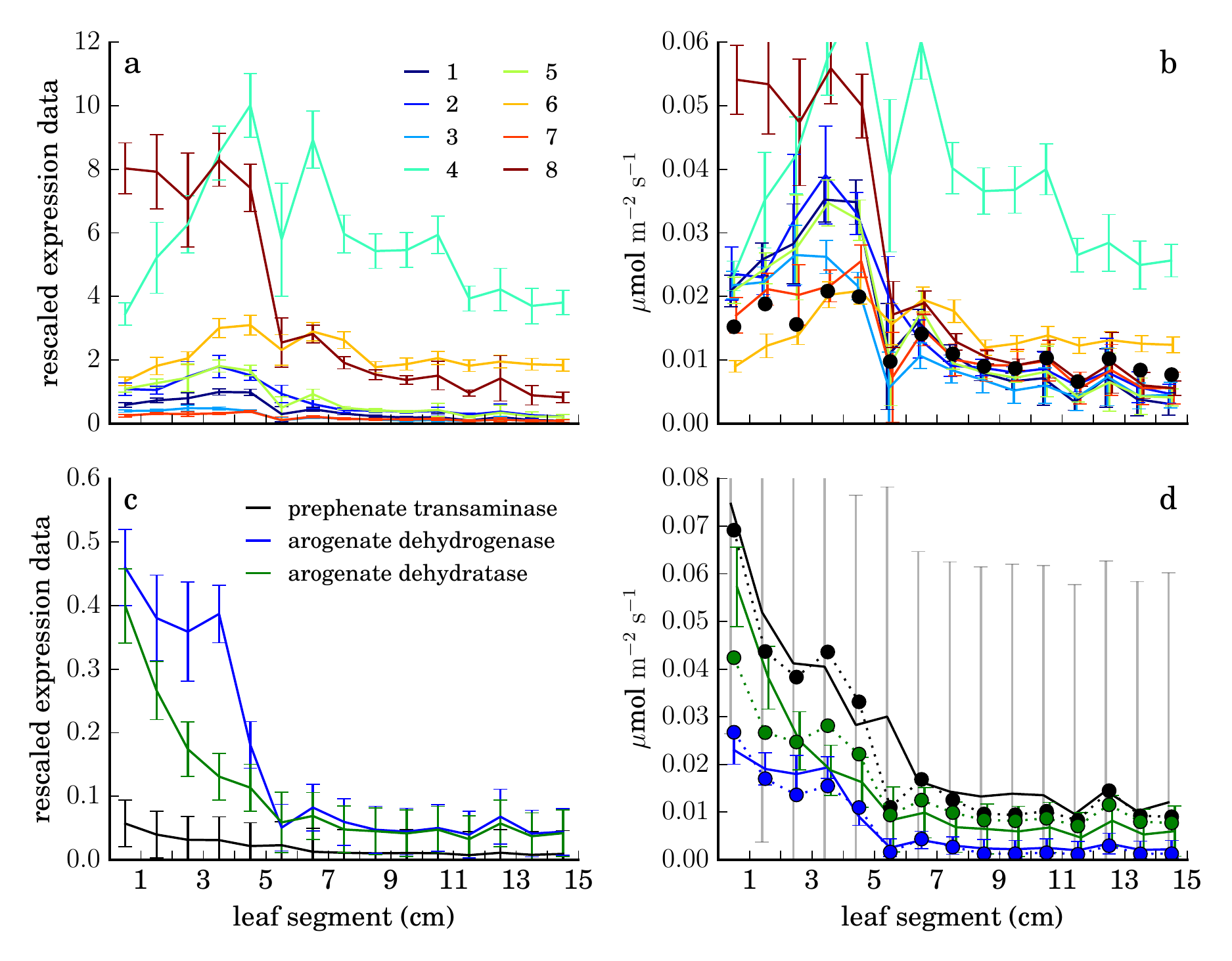}
  \caption{{\bf Comparison of RNA-seq data to predicted fluxes for a
      linear pathway and around a metabolic branch point.} Upper
    panels, chlorophyllide a synthesis in the mesophyll; lower panels,
    production of arogenate in the bundle sheath by prephenate
    transaminase and its consumption by arogenate dehydrogenase and
    arogenate dehydratase.  Left, aggregate RNA-seq data and
    experimental standard deviations for each reaction rescaled by a
    uniform factor (see text).  Right, same data and errors further
    rescaled by reaction-specific optimal factors ($e^{-s_i}$, in the
    variables of eq. \ref{fitting}) to best match data with predicted
    fluxes (solid circles).  Fluxes are equal for all reactions of the
    linear pathway (1, uroporphyrinogen decarboxylase, 2,
    coproporphyrinogen oxidase, 3, protoporphyrinogen oxidase, 4,
    magnesium chelatase, 5, magnesium protoporphyrin IX
    methyltransferase, 6, magnesium protoporphyrin IX monomethyl ester
    cyclase, 7, divinyl chlorophyllide a 8-vinyl-reductase, 8,
    protochlorophyllide reductase.) Error bars represent standard
    deviations of expression measurements across multiple replicates.}
\label{pathwayfig}
\vskip 5.0cm
\end{figure}

Figure~\ref{pathwayfig} illustrates the operation of the fitting
algorithm in detail, using two regions of the metabolic network with
simple structure as examples.

In Fig.~\ref{pathwayfig}a, expression data for eight reactions of
the pathway leading to chlorophyllide a are shown. Expression levels
for the different reactions at any point on the leaf may span an order
of magnitude or more, but the FBA steady-state assumption requires the
rates of all reactions in this unbranched\footnote{The branch leading
  to heme production is not included in the reconstruction.}  pathway
to be equal at each point. Applying the optimal rescaling determined
for each reaction's expression data, shown in panel b, allows the flux
prediction for the pathway (solid dots) to achieve reasonable
agreement with the data. (Note that data for reaction 4 cannot be
further scaled down because of the lower limit $\exp (-5)$ on its
scale factor $\exp (s_4)$, imposed for technical reasons.)

Figure \ref{pathwayfig}c shows data for a three-reaction branch point
in aromatic amino acid synthesis. To balance production and
consumption of arogenate, the prephenate transaminase flux must equal
the sum of the fluxes through arogenate dehydrogenase (to tyrosine)
and arogenate dehydratase (to phenylalanine) but expression is
consistently lower for the transaminase than the other enzymes. After
rescaling (Fig.~\ref{pathwayfig}d), the data agree well with the
stoichiometrically consistent flux predictions (solid dots). The
predicted ratio of dehydrogenase to dehydratase flux reflects data for
downstream reactions.

\subsubsection*{Comparison to other methods for integrating RNA-seq data}

\autoref{S4_Figure} shows predictions that result when the scale factors $s_i$
of eq. (\ref{fitting}) are fixed to zero. The source-sink transition
is apparent but the C4 cycle operates at lower levels, the example
pathways of Fig.~\ref{pathwayfig} (and a number of others) show little
or no activity, and predicted fluxes along the leaf are not as tightly
correlated with their associated expression data.

\autoref{S5_Figure} shows the metabolic state predicted by applying the
expression data for each reaction as an upper bound on the absolute
value of the reaction rate as in the E-Flux method~\cite{Colijn2009}
to the fifteen-segment model with the same RNA-seq data.  The C4
system is predicted to operate, but no source-sink transition is
apparent, and typical data-predicted flux correlations are
poor. Imposing a realistic biomass composition restores the
source-sink transition and somewhat improves correlation between data
and fluxes (\autoref{S6_Figure}). Fluxes predicted by E-Flux are generally
smaller than those predicted by the least-squares method, with or
without per-reaction scale factors.

\autoref{S18_Figure} compares the fluxes predicted at the tip by
optimizing agreement with the data through the non-biological
objective function (eq.~\ref{fitting}), fluxes predicted at the tip
with an explicit biological objective function (maximizing
CO\textsubscript{2} assimilation) constrained by the experimental data
in the E-Flux method, and fluxes predicted in an FBA calculation which
ignores the data entirely (minimizing total flux while achieving the
same CO\textsubscript{2} assimilation rate as predicted at the tip by
the least-squares method.) Both data-integration methods lead to predictions
very different from the unconstrained FBA calculation.

\section*{Discussion}
\subsection*{Reconstruction}
Our model is the fourth published genome-scale metabolic
reconstruction of the major crop plant \textit{Zea mays}, and the
first such reconstruction developed solely from maize data sources,
rather than as a direct or indirect adaptation of the
\textit{Arabidopsis thaliana} model AraGEM
\cite{deOliveiraDalMolin2010}.

Direct reaction-to-reaction comparison of iEB5204 with C4GEM
\cite{GomesdeOliveiraDalMolin2010}, iRS1563~\cite{Saha2011}, and its
successor model~\cite{Simons2014a} is difficult because those models
use a naming scheme for compounds and reactions ultimately based on
KEGG~\cite{Kanehisa2014,Kanehisa2000} while this model, like its
parent database, uses the nomenclature of MetaCyc and the BioCyc
database collection. The models are broadly similar in size and
biological scope. As published, C4GEM included {1588} reactions
associated with {11623} maize genes; iRS1563, {1985} reactions
associated with {1563} genes; the model of Simons et
al.~\cite{Simons2014a}, {3892} unique reactions and {5824} genes; and
iEB5204, {2720} reactions with {5204} genes. All models can simulate
the production of similar sets of basic biomass constituents
(including amino acids, carbohydrates, nucleic acids, lipids and fatty
acids, and cell wall components) under photosynthetic and
non-photosynthetic conditions and include key reactions of the C4
cycle. The model of Simons et al.~\cite{Simons2014a} also offers
extensive coverage of secondary metabolism.

However, the present model has several advantages which make it
particularly suitable for integration with transcriptomics data:
\begin{description}
\item[Gene associations] The gene associations included in iEB5204 are
  those presented in CornCyc~\cite{CornCyc}, which are generated by
  the PMN Ensemble Enzyme Prediction Pipeline (E2P2)\cite{E2P2}, a
  homology-based protein sequence annotation algorithm trained on a
  reference dataset of experimentally validated enzyme sequences.  The
  E2P2 approach is more comprehensive and scalable than the
  development procedures of the previous maize reconstructions (which
  involve, for example, obtaining gene associations by transferring
  annotations from Arabidopsis genes to their best maize BLAST hits
  and manually selecting annotations for remaining maize genes from
  among BLAST hits in other species.)  The entire set of gene
  associations in the FBA model may be readily updated based on
  improvements in the E2P2 prediction algorithm.

\item[High-confidence submodel] In developing the fitting algorithm we
  found that, to obtain plausible metabolic state predictions, a
  conservative reconstruction was preferable to a comprehensive one.
  For example, early tests with the comprehensive version of the model
  suggested that the fitting algorithm often found low-cost solutions
  involving high fluxes through reactions which, on investigation, we
  determined were unlikely to be active in maize.  Because of the
  model's connection to the CornCyc database, it was straightforward to
  create a reduced, high-confidence version of the model by
  preferentially excluding reactions not included in any manually
  curated plant metabolic pathway, even if candidate associated genes
  had been identified computationally, leading to more realistic
  results.

\item[Reproducibility] 
  In an effort to improve the reusability of the model and encourage
  its application to other data sets, we have provided the full source
  code (\nameref{S14_Protocol} and \nameref{S15_Protocol}) for all
  calculations presented here, as has been recommended (see,
  e.g.,~\cite{Sandve2013}).
\end{description}

Previous reconstructions do offer two features absent from this
model: gene associations for intracellular transport reactions, and
gene associations which take into account the structure of protein
complexes.  Both should be considered in future work.

In agreement with \cite{Latendresse2012}, we found that building the
model starting from a metabolic pathway database was considerably more
straightforward than the standard process of \textit{de novo}
reconstruction~\cite{Thiele2010}.  Reasonable effort was still
required to bring the model to a functional state by identifying
reactions or pathways present in the CornCyc database which could not
be handled automatically by the Pathway Tools export facility (for
example, because they involved polymerization, or could not be checked
automatically for conservation violations) and determining how to
represent them appropriately in the FBA model.

The model construction process here could readily be adapted to
generate metabolic models describing any of the more than 30 crop and
model plant species for which Pathway Tools-based metabolic pathway
databases \cite{Karp2010} have been developed by 
the Plant Metabolic Network \cite{PMNOverview}, 
Sol Genomics Network \cite{Fernandez-Pozo2014}, 
Gramene \cite{Monaco2014}, 
and others (e.g.,\cite{Urbanczyk-Wochniak2007,Naithani2014,Jung2014}) 
allowing the present data-fitting method to be applied to RNA-seq data
from those organisms. The level of model development effort required
and quality of fit results will vary depending on the extent of
curation of the pathway database and quality of the gene function
annotations.


\subsection*{Nonlinear optimization}
In contrast to the linear and convex optimization methods employed in
nearly all prior constraint-based modeling work, general constrained
nonlinear optimization algorithms typically require more effort from
the user (who might be required to supply functions which evaluate the
first and second derivatives of all constraints with respect to all
variables in the problem). They are slower, are more sensitive to
choices of starting point and problem formulation, are not guaranteed
to converge to an optimal point even if one exists, and, when they do
converge to an optimum, cannot guarantee that it is globally optimal.

The software package we present allows the rapid and effective
development of metabolic models with nonlinear constraints despite
these complications. All necessary derivatives of constraint functions
are taken analytically, and Python code to evaluate them is
automatically generated. A model in SBML format may be imported,
nonlinear constraints added and removed, and the problem repeatedly
solved to test various design choices, solver options, and initial
points, all within an interactive session, with a minimum of initial
investment of effort in programming.

In the present case, agreement between nonlinear FBA calculations
maximizing growth and the predictions of classical physiological
models confirmed that the true, globally optimal CO\textsubscript{2}
assimilation rate was found successfully.  For the data-fitting
calculations, where the true optimal cost is not known, we cannot
exclude the possibility that there exist other optimal solutions,
qualitatively distinct from the flux distributions and quasi-optimal
regions presented above, with equivalent or lower costs. In practice,
we encountered occasional cases in which reaction or pathway fluxes
were initially predicted to be zero even when associated with nonzero
data, despite the existence of a superior alternative solution with
nonzero predicted fluxes. A step to detect and correct these
situations was incorporated into the fitting algorithm.

Many future applications for the software are possible.  Our approach
to Rubisco kinetics may easily be extended to other models of C4
metabolism or, more generally, to any FBA calculation in a
photosynthetic organism where the CO\textsubscript{2} level at the Rubisco
active site, and thus the Rubisco oxygenation/carboxylation ratio, is
not known \textit{a priori}.  A recent genome-scale metabolic
reconstruction of the model alga \textit{Chlamydomonas reinhardtii},
for example, was identified by the authors as being deficient in
describing algal metabolism under low CO\textsubscript{2} conditions due to
the fact that the Rubisco carboxylase and oxygenase fluxes were
treated as independent and not competitive, as we have done
here~\cite{Chang2011}.

Ensuring that rates of Rubisco oxygenation, Rubisco carboxylation, and
PEPC carboxylation are consistent with our knowledge of their kinetics
is a special case of the more general problem of integrating kinetic
and constraint-based modeling, to which diverse approaches have been
proposed (e.g., \cite{Mahadevan2002,Smallbone2007,
  Jamshidi2010,Feng2012,Cotten2013,Chowdhury2014}).

To our knowledge, no prior work has simply imposed kinetic laws as
additional, nonlinear constraints in the ordinary FBA optimization
problem. Our results demonstrate the potential of this approach in
systems where the kinetics of a few well-understood reactions are
crucial. It remains to be seen how many kinetic laws may be
incorporated in this way at once, and to what extent their
introduction usefully constrains the space of possible steady-state
flux distributions even when relevant kinetic parameters are not known
(but instead are treated as optimizable variables, an approach with
connections to ensemble kinetic modeling~\cite{Tan2011}).
 
Nonlinear constraints may also be of use in enforcing thermodynamic
realizability of flux distributions, and relaxing requirements of
linearity or convexity may stimulate the development of novel
objective functions -- either for data integration purposes, as here,
or as alternatives to growth-rate maximization.

\subsection*{Data fitting}
The expression of a gene encoding a metabolic enzyme need not
correlate with the rate of the reaction that enzyme catalyzes. The
relationship between transcription and degradation of mRNA and control
of flux is indirect, mediated by protein translation, folding, and
degradation, complex formation, posttranslational modification,
allosteric regulation, and substrate availability. Indeed,  as reviewed
by \cite{Hoppe2012}, experimentally observed correlations among
RNA-seq or microarray data (each itself an imperfect proxy for mRNA
abundance or transcription rate), protein abundance, enzyme activity,
and fluxes are variable and often weak.

For example, RNA-seq and quantitative proteomic data obtained from
maize leaves at the same developmental stage studied here, harvested
simultaneously from plants grown together, showed Pearson correlation
approximately 0.6 across the entire dataset, but some significantly
lower values were found when correlations were restricted to genes of
particular functional classes, and measured mRNA/protein ratios for
individual genes varied up to 10-fold along the gradient
\cite{Ponnala2014}. A subset of this data is shown in
Fig.~\ref{costfig}d.

The most comprehensive study of the issue in plants so far
\cite{temp_schwender} found so little agreement between RNA-seq and
13C-MFA data from embryos of two \textit{Brassica napus} accessions
that the authors concluded the inference of central metabolic fluxes
from transcriptomics is, in general, impossible.

In this light, it is not surprising that methods for integrating
transcriptomic data with metabolic models to predict reaction rates
have met with limited success. Machado and Herrgård
\cite{Machado2014a} reviewed 18 such methods and assessed the
performance of seven of them on three test datasets from
\textit{E. coli} and \textit{Saccharomyces cerevisiae} where
experimentally measured intracellular and extracellular fluxes were
available for comparison. None of the methods consistently
outperformed parsimonious FBA simulations which completely ignored
transcriptomic data.

In contrast, in the present work the use of transcriptomic data (and a
limited number of enzyme activity measurements) allowed the correct
prediction of a metabolic transition from the base of the leaf to the
tip, which could not have been expected based on FBA calculations
alone: without such data, all points along the gradient would be
identical, and the biomass-production-maximizing solution would be the
same at each. The predicted position of the source-sink transition is
not perfectly accurate, and the overall performance of the model
cannot be evaluated until the predicted reaction rates are compared to
detailed experimental flux measurements. Nonetheless, the results are
encouraging. We offer two explanations for this apparent success.

First, the metabolic transition between the heterotrophic sink region
at the base and the photoautotrophic source region at the tip is
particularly dramatic, involving a large number of reactions which are
effectively absent in one region but carry high fluxes in the other~\cite{Li2010};
so long as even a slight correlation between transcript levels and
fluxes exists, such a reconfiguration should be apparent from
expression data.

Second, although the developing maize leaf is biologically more
complex than microbial growth experiments, the relationship between
expression levels and fluxes may be actually be closer in the
leaf. Leaf development is a stereotyped, frequently repeated,
relatively slow, one-way process, in which the precise sequence of
events is subject to evolutionary optimization.  Coordination of
transcription with required fluxes will lead to efficient use of
resources. In contrast, the test cases of \cite{Machado2014a} involve
microbial responses to varying environmental conditions and under- and
over-expression mutations. Environmental responses must be rapid,
flexible and reversible -- criteria a complex, scripted transcriptional
response may not satisfy -- while transcriptional responses to novel
mutations, by definition, cannot have been evolutionarily optimized.
This hypothesis could be tested by evaluating performance of the
present method on RNA-seq data from mutant maize plants, or plants
subject to environmental challenges.

We note also that methods that did not constrain or optimize the
growth rate predicted zero growth rates in almost all the test cases
studied by Machado and Herrgård \cite{Machado2014a}. The present
method also does not constrain or optimize the growth rate but
consistently does predict nonzero growth as reflected in nonzero biomass
production (whether with a flexible biomass
composition was used, as above, or a fixed biomass composition, as in
\autoref{S7_Figure} and \autoref{S8_Figure}).

\subsection*{The whole-leaf model}
Large-scale metabolic models of interacting cells of multiple types
first appeared in 2010, with C4GEM~\cite{GomesdeOliveiraDalMolin2010}
and a model of human neurons interacting with their surrounding
astrocytes~\cite{Lewis2010}.  Many more complex multicellular FBA
models have since appeared, including studies of the metabolism of
interacting communities of microbial species in diverse natural
environments or artificial
co-cultures~\cite{Salimi2010,Zhuang2011,Zomorrodi2012,Zengler2012,
  Khandelwal2013,Chiu2014, Zomorrodi2014} (also~\cite{Stolyar2007} at
a smaller scale) and of the metabolic capacities of host animals and
their symbionts~\cite{Bordbar2010} or parasites~\cite{Heinken2013}.
In plants, diurnal variation in C3 and CAM plant metabolism has been
simulated with a model which represents different phases of the
diurnal cycle with different abstract compartments, with transport
reactions representing accumulation of metabolites over
time~\cite{Cheung2014}.

In the most direct antecedent of the present work, Grafahrend-Belau
and coauthors developed a multiscale model of barley
metabolism~\cite{Grafahrend-Belau2013} which represented leaf, stem,
and seed organs as subcompartments of a whole-plant FBA model, with
nutrients exchanged through the phloem. Combining the FBA model with a
high-level dynamic model of plant metabolism allowed them to predict
changes in metabolism over time, including the transition between a
biomass-producing sink state and a fructan-remobilizing source state
in the stem late in the plant's life cycle.

The whole-leaf model presented here occupies an intermediate position
between prior C4 models, with single mesophyll and bundle sheath
cells, and multi-organ whole-plant models such
as~\cite{Grafahrend-Belau2013}.  It represents the first attempt to
model spatial variations in metabolic state within a single organ,
allowing the study of developmental transitions in leaf metabolism by
incorporating data from more and less differentiated cells at a single
point in time, rather than modeling development dynamically.

Other interacting cell models incorporate \textit{a priori}
qualitative differences in the metabolic capabilities of their
components (e.g., leaf, stem, and seed, or neurons and astrocytes).
In contrast in the work presented here, in order to allow the
metabolic differences between any two adjacent points to be purely
quantitative, the same metabolic network must be used for all
points. This simplifies the process of model creation but implies that
meaningful predictions of spatial variation depend entirely on the
integration of (spatially resolved) experimental data. The ability of
the model to capture the experimentally observed shift from sink to
source tissue along the developmental gradient based on RNA-seq and
enzyme activity measurements shows that this may be done successfully
with high-resolution -omics data and careful model construction.

\section*{Methods}
\subsection*{Reconstruction process}
A local copy of CornCyc 4.0 \cite{CornCyc} was obtained from the
Plant Metabolic Network and a draft metabolic model was created using
the MetaFlux module of Pathway Tools 17.0~\cite{Latendresse2012}. The
resulting model, including reaction reversibility information, was
converted to SBML format and iteratively revised, as described in
detail in \nameref{S9_Appendix}, until all desired biomass components
could be produced under both heterotrophic and photosynthetic
conditions and realistic mitochondrial respiration and
photorespiration could
operate. 

An overall biomass reaction was adapted from iRS1563~\cite{Saha2011}
with minor modifications to components and stoichiometry, as detailed
in \nameref{S9_Appendix}. To allow calculations with flexible
biomass composition, individual sink reactions were added for most
species participating in the biomass reaction, as well as several
relevant species (including chlorophyll) not originally included in
the iRS1563 biomass equation.

Core metabolic pathways were assigned appropriately to subcellular
compartments (e.g., the TCA cycle and mitochondrial electron transport
chain to the mitochondrion; the light reactions of photosynthesis, the
Calvin cycle, and some reactions of the C4 cycle to the chloroplast;
and some reactions of the photorespiratory pathway to the peroxisome)
and the intracellular transport reactions necessary for their
operation were added.

The model was thoroughly tested for consistency and conservation
violations, confirming that no species could be created without net
mass input or destroyed without net mass output (except species
representing light, which can be consumed to drive futile cycles.)

The base metabolic model iEB5204 is provided in SBML format as
\nameref{S11_Model}. Gene association rules for reactions with
associated genes in CornCyc are provided following COBRA
conventions~\cite{Becker2007}.  Additional annotations give the record
in the CornCyc database associated with each reaction and species,
where applicable.

To produce the higher-confidence version of the reconstruction,
iEB2140 (\nameref{S12_Model}), reactions in the base model which were
not associated with any identified metabolic pathway in CornCyc, and
those for which no genes for a catalyzing enzyme had been identified
by computational function prediction, were removed from the model if
their removal did not prevent photosynthesis, photorespiration, or the
production of any biomass component.  Then, all reactions which could
not achieve nonzero steady-state rates were removed.

\subsection*{Mesophyll-bundle sheath model}
A model for leaf tissue (\nameref{S13_Model}) was created by taking
two copies of the high-confidence model, representing mesophyll and
bundle sheath cells, and adding reactions representing transport
through the plasmodesmata which connect the cytoplasmic spaces of
adjacent cells. Though in principle most small molecules can cross the
plasmodesmata by diffusion~\cite{Weiner1988}, unrealistic
concentration gradients may be required to drive high diffusive
fluxes, and processes other than simple diffusion may play a role in
the rapid exchanges which do occur~\cite{Sowinski2008}. Given this
uncertainty we conservatively restricted such transport to species
known or expected to be exchanged between cell types (under at least
some circumstances); a complete list is given in \nameref{S9_Appendix}.

Net import or export of metabolites from the system was limited to the
mesophyll, for gases exchanged with the intercellular airspace, or the
bundle sheath, for soluble metabolites exchanged with the leaf's
vascular system. Reactions were not otherwise restricted \textit{a
  priori} to a particular cell type. To facilitate integration with
cell-type-specific RNA data, gene associations in this model are
tagged with the relevant cell type, e.g. `bs\_GRMZM2G039273' vs
`ms\_GRMZM2G039273'.

\subsection*{Leaf gradient model} 
The choice of phloem transport metabolites (other than sucrose) is a
compromise. Glycine is the most abundant amino acid in maize
phloem~\cite{Ohshima1990}, and glutathione is a putative phloem sulfur
transport compound~\cite{Bourgis1999}, but many other amino acids are
present in the phloem sap, and other compounds (e.g.,
S-methyl-methionine~\cite{Bourgis1999}) may play roles in phloem
sulfur transport. However, we found that the available data did not
adequately constrain rates of phloem transport if multiple transport
species of each type were allowed, resulting in high rates of
transport from the base towards the tip, against the direction of bulk
flow in the phloem.

For simplicity, export of metabolites from the leaf to the rest of the
plant through the phloem was neglected and net import of sucrose was
not allowed.  Each segment was taken to have the same total area, so
that a 1 \fluxunit{} rate of sucrose loading
in one segment exactly balanced a 1 \fluxunit{}
rate of sucrose unloading in another segment.

Note that the whole-leaf model is constructed dynamically within the
data-fitting code, rather than being loaded from an SBML file.

\subsection*{Physiological constraints}
Rubisco carboxylase and oxygenase rates $v_c$ and $v_o$ in mesophyll and 
bundle sheath chloroplasts were constrained to obey Michaelis-Menten
kinetic laws with competitive inhibition,
\begin{equation}
  \label{rubisco_kinetics}
  \begin{aligned}
    v_c &=\frac {v_{c,\text{max}} \left[\text{CO}_2\right]}
              {\left[\text{CO}_2\right] + k_c \left( 1 + \frac{\left[\text{O}_2\right]}{k_o}\right)}\\
    v_o &=\frac {v_{o,\text{max}} \left[\text{O}_2\right]}
              {\left[\text{O}_2\right] + k_o \left( 1 + \frac{\left[\text{CO}_2\right]}{k_c}\right)},\\
  \end{aligned}
\end{equation}
and the relationship $v_{o,\text{max}}/v_{c,\text{max}}= k_C/(k_O\cdot
S_R)$ was imposed, from which eq. (\ref{vo_vc_ratio})
follows~\cite{VonCaemmerer2000}. The Michaelis-Menten constants for
oxygen and carbon dioxide $k_C$ and $k_O$ and the Rubisco specificity
$S_R$ were set to values typical of C4 species: $k_C$, 650 \textmu
mol\,mol\textsuperscript{-1}; $k_O$, 450
mmol\,mol\textsuperscript{-1}; $S_R$, {2590}~\cite{VonCaemmerer2000}.

The rate of PEP carboxylation in the mesophyll was bounded above by an
appropriate kinetic law,
\begin{equation}\label{pepc_kinetics}
v_p = \frac {v_{p,\text{max}} \left[\text{CO}_2\right]}
              {k_{C,p} + \left[\text{CO}_2\right]}
\end{equation}
with $0 \leq v_{p,\text{active}} \leq v_{p,\text{max}} $ and an
appropriate $k_{C,p}$ ({80}mmol\,mol\textsuperscript{-1},
\cite{VonCaemmerer2000}).

The parameters $v_{p\text{max}}$ and $v_{c,\text{max}}$ representing
the total amount of Rubisco and PEPC available may be fixed to permit
comparison to models parameterized in those terms or allowed to vary.

Rates of oxygen and carbon dioxide diffusion from the bundle sheath to
the mesophyll, $L$ and $L_O$, were constrained to obey the relationship
\begin{equation}
  \label{leakiness}
  \begin{aligned}
    L &= g_{BS}\left(\text{CO}_{2,BS} - \text{CO}_{2,ME}\right) \\
    L_O &= g_{BS,O}\left(\text{O}_{2,BS} - \text{O}_{2,ME}\right)
  \end{aligned}
\end{equation}
with $g_{BS,O}=0.047g_{BS}$ \cite{VonCaemmerer2000}. All simulations
used the bundle sheath CO\textsubscript{2} conductivity measured by
\cite{Bellasio2014} for maize plants grown under high light, $1.03 \pm
0.18$ \fluxunit{}. While $g_{BS}$ undoubtedly varies along the
developmental gradient, its deviation from this value (measured in
fully-expanded leaves, 3-4 weeks after planting) is likely greatest
below the region of high suberin synthesis identified 4 cm from the
leaf base \cite{fifteensegment}; as the C4 cycle was not predicted to
operate at high rates in this region, the impact of this discrepancy
should be limited.

Resistance to CO\textsubscript{2} diffusion from the intercellular
airspace to the mesophyll cells was neglected;
ref.~\cite{Kromdijk2010} reported $g_m\approx 1$
mmol\,m\textsuperscript{-2}\,s\textsuperscript{-1} in maize under a
variety of conditions, suggesting the mesophyll and intercellular
CO\textsubscript{2} levels would differ only slightly at the rates of
CO\textsubscript{2} assimilation and release dealt with
here. Similarly, all intracellular compartments were taken to have
equal CO\textsubscript{2} concentrations.

\subsection*{Optimization calculations}
The nonlinear modeling package uses the libsbml python bindings to
read SBML files \cite{Bornstein2008} and an internal representation of
SBML models derived from the SloppyCell package \cite{SloppyCell,
  Myers2007}. IPOPT calculations used version 3.11.8 with the linear
solver ma97 from the HSL Mathematical Software Library
\cite{HSL}. Where not specified, convergence tolerance was $10^{-5}$,
or $10^{-4}$ in FVA calculations.  To solve purely linear problems
(e.g., to test the production of biomass species during the
reconstruction process, where nonlinear constraints were not used) the
GNU Linear Programming Kit, version 4.47 \cite{glpk}, was called
through a Python interface \cite{pyglpk}.

\subsection*{Comparison with other models}
Python code used to calculate the predictions of the models of
von Caemmerer~\cite{VonCaemmerer2000} for comparison with nonlinear optimization
results is provided in \nameref{S14_Protocol}.

\subsection*{Integrating biochemical and RNA-seq data}
\subsubsection*{RNA-seq datasets}
\label{rnadatasets}
To obtain mesophyll- and bundle-sheath-specific expression levels at
15 points, we combined the non-tissue-type-specific data of Wang et
al.~\cite{fifteensegment}, measured at 1-cm spatial resolution, with
the tissue-specific data of Tausta et al.~\cite{Tausta2014} obtained
by using laser capture microdissection (LCM) -- measured 4 cm, 8 cm
and 13 cm from the leaf base (the upper three highlighted positions in
Fig.~\ref{sourcesinkfig}b).  This integration was achieved by
determining for each gene at each of those points with LCM data the
ratio of the average RPKM in the mesophyll ($M$) to the sum of the
average RPKM values for mesophyll and bundle sheath ($M+B$);
furthermore, we assumed that the $M/(M+B)$ ratio at the leaf base was
0.5 (based on the proteomic experiments of Majeran et
al.~\cite{Majeran2010}, which showed only limited mesophyll-bundle
sheath specificity there), and linearly interpolating to estimate
$M/(M+B)$ ratios at all 15 points. For very weakly expressed genes, we
did not impose cell-type specificity: where the sum of mesophyll and
bundle sheath RPKM in the LCM data was less than 0.1, we assumed
$M/(M+B)=0.5$. We then divided the mean whole-leaf FPKM measurement at
each point into mesophyll and bundle sheath portions according to
these ratios.

To associate expression data with a reaction, data for its associated
genes were summed, dividing the data for a gene associated with
multiple reactions in the model equally among them.  The uncertainties
$\delta_{ij}$ in the objective function (eq.~(\ref{fitting})) were
estimated in an ad hoc way by splitting the standard deviations of the
FPKM values over multiple experimental replicates according to the
$M/(M+B)$ ratios and then summing the uncertainties for all genes
associated with a particular reaction, imposing a minimum relative
error of 0.05 and a minimum absolute uncertainty corresponding to 7.5
FPKM.

To globally rescale the expression data to be comparable to expected
flux values, data for PEPC and Rubisco were compared to the enzyme
activity measurements discussed below and a simple linear regression
performed, yielding a conversion factor of 204 FPKM $\approx$ 1
\fluxunit{} for these enzymes. All expression data were divided by
this factor before solving the optimization problem.

\subsubsection*{Enzyme activity measurements}
Enzyme activities constrained by measurements in \cite{fifteensegment} 
were alanine aminotransferse, aspartate aminotransferase,
fructose bisphosphate aldolase, 
glyceraldehyde 3-phosphate dehydrogenase (NADPH), 
glyceraldehyde 3-phosphate dehydrogenase (NADH),
glutamate dehydrogenase (NADH),
malate dehydrogenase (NADH),
malate dehydrogenase (NADPH),
PEPC,
phosphofructokinase,
phosphoglucomutase,
phosphoglucose isomerase, 
phosphoglycerokinase,
Rubisco,
transketolase,
triose phosphate isomerase,
and UDP-glucose pyrophosphorylase.

For Rubisco and PEPC, enzyme data constrained the sum of the variable
kinetic parameters $v_{c,\text{max}}$ and $v_{p,\text{max}}$ in
mesophyll and bundle sheath compartments, rather than the sum of the
associated fluxes. Enzyme data in nanomole per minute per gram fresh
weight was converted to micromole per second per square meter of leaf
surface area assuming a fresh weight of 150 g\,m\textsuperscript{-2}.

\subsubsection*{Handling reversible reactions}
The objective function (eq. (\ref{fitting})) optimizes the agreement
between the absolute value of the flux through each reaction with its
data, but IPOPT requires a twice continuously differentiable objective
function.  We use a reformulation $F'$ representing each absolute value
$|v_{ij}|$ as the product of the flux and a parameter $\sigma_{ij}$
representing its sign:
\begin{equation}\label{fitting_no_abs}
  F'(v) = \sum_{i=0}^{N_r}\sum_{j=1}^{15} \frac{\left(e^{s_i}\sigma_{ij}v_{ij}-d_{ij}\right)^2}
  {\delta^2_{ij}} + 
  \alpha \sum_{i=0}^{N_r} s_i^2
\end{equation}
Similarly, the enzyme activity data constraint,
eq. (\ref{enzymeequation}), was rewritten to replace absolute values in
this way. Reaction rates with positive (negative) sign parameter were
required to take values greater than a small negative (less than a
small positive) tolerance, typically 1.0.

Choosing the $\sigma_{ij}$ to optimize $F'$ is a very large scale
mixed-integer nonlinear programming problem. We arrive at an
approximate solution using a heuristic method similar in spirit to
that of \cite{Lee2012}, with three steps.
\begin{enumerate}
\item The subproblems representing each segment of the leaf are solved
  separately, with all scales $s_{i}$ set to zero and modest upper
  and lower bounds on the reactions representing nutrient exchange
  with the phloem. Within each segment, a sign for the reversible
  reaction $r_1$ with the highest associated expression data is chosen
  by first setting its sign $\sigma_1$ to $+1$, finding the
  minimum-flux best-fitting flux distribution $\mathbf{v}^+$ ignoring
  the costs associated with all other reversible reactions (but
  including costs associated with all irreversible reactions), then
  finding the cost $c^+$ of the best-fitting flux distribution
  $\mathbf{v'}^+$ considering the costs of the reversible reactions
  with nonzero fluxes in $\mathbf{v}^+$ (temporarily setting their
  signs according to their values in that case.) A cost $c^{-}$ is
  determined analogously after setting the sign $\sigma_1$ to $-1$ ,
  and if $c^{-}<c^{+}$, $\sigma_1=-1$ is chosen; otherwise,
  $\sigma_1=+1$.  Then the reversible reaction with the second-highest
  expression data $r_2$ is treated in the same way, considering $r_1$
  to be irreversible.

\item When signs for all reversible reactions have been chosen at a
  segment, a final best-fitting flux distribution given those signs is
  determined.  Then the full optimization problem, combining all
  fifteen segments, is solved with the chosen sign parameters fixed,
  using those flux distributions to provide a nearly-feasible initial
  guess.  

\item The sign-choice process in each subproblem is then solved again,
  fixing the scale factors $s_{i}$ and rates of metabolite exchange
  with the phloem to those determined in the full problem.  If no
  signs change, or if the new signs do not decrease the objective function
  value, fitting stops; otherwise, step 2 is repeated.

\item Finally, for each reaction $j$ with nonzero data and maximum
  absolute flux less than ${0.0001}$ at any point in the leaf
  model, a lower bound of $-0.99d_i$ is imposed on the term
  $\left(e^{s_i}\sigma_{ij}v_{ij}-d_{ij}\right)$ in the objective
  function, for $i=1,\ldots,15$, and the full fifteen-segment optimization 
  problem is solved again. 
\end{enumerate}
The final step addresses the observation that the optimization process
occasionally converged to a solution in which a few reactions with
associated data were predicted to have zero flux when a better solution
with nonzero flux existed. In some cases (e.g. the $s_i=0$ case shown in
\autoref{S4_Figure}) this step did not lead to an overall reduction in
the objective function and was omitted.

Steps 1 and 3 take between one and eight hours per segment using an AMD
Opteron 6272 and may be easily parallelized across up to 15
processors. Step 2 may take up to 2 hours in the first iteration but
is often faster in later iterations, when the initial guess is
closer to the optimum. Typically the procedure stops after 4-5
iterations, requiring about 24 total hours of wall time using 15
processors.

\subsubsection*{Special cases}
The Rubisco oxygenase, Rubisco carboxylase, and mesophyll PEPC fluxes
are excluded from the objective function.  Instead, terms are added
comparing the transcriptomic data for those enzymes to the variables
which explicitly represent their activity level: for Rubisco,
$v_{c,\text{max}}$ in mesophyll and bundle sheath compartments, and
for PEPC, $v_{p\text{max}}$ in the mesophyll. Scale factors for the
mesophyll and bundle sheath Rubisco activities are not constrained to 
be equal.

\section*{Acknowledgments}
This work was supported by National Science Foundation grant
IOS-1127017 and a grant to the International Rice Research Institute
from the Bill and Melinda Gates Foundation.  The authors thank Tom
Brutnell, Lin Wang, Lori Tausta, Qi Sun, Zehong Ding, and Tim Nelson
for data and comments, Sue Rhee, Kate Dreher, and Peifen Zhang for
helpful discussion of our use of CornCyc, and Lei Huang and Brandon
Barker for discussions of metabolic modeling.

\bibliographystyle{plos2009}
\bibliography{multiscale_c4}

\addresseshere
\clearpage
\section*{Supporting Information}
\section{Outline}
\label{S9_Appendix} 
This appendix describes the of creation of a metabolic model for maize
from CornCyc.  It covers the creation of an SBML model with exchange
and biomass reactions and limited subcellular compartmentalization
which can successfully simulate the production of many biomass
components and photosynthetic carbon dioxide assimilation, the
adaptation of the biomass equation from iRS1563, some considerations
in the process of expanding the model to describe interacting
mesophyll and bundle sheath compartments, and some modifications made
in response to preliminary fitting results.

Sections \ref{export} through \ref{sbml} explain in detail the process
of constructing the underlying metabolic model at the one-cell
level. Section \ref{refinement} discusses in detail changes made to
gene associations based on early data fitting results.  Section
\ref{biomass} describes changes to the iRS1563 biomass equation.
Section \ref{plasmodesmata} discusses plasmodesmatal transport in the
two-cell model.  Filenames referred to are in the
\texttt{model\_development} subdirectory of the project source code
(see S15 Protocol.)

\section{Exporting the CornCyc FBA model from Pathway Tools}
\label{export}
CornCyc 4.0 \cite{CornCyc} was obtained from the Plant Metabolic Network
and upgraded from from Pathway Tools 16.5 to 17.0 locally.

The frame \texttt{PWY-561} was removed from the database because
otherwise some of the reactions of that pathway were excluded from the
FBA export, apparently due to a bug.

A simple FBA problem was solved using the Pathway Tools FBA
functionality \cite{PathwayToolsManual}, producing an output file
which includes all reactions in the FBA model Pathway Tools generates
internally, both those which are active in the solution to the FBA
problem and those which are not. Note that this list of reactions is
distinct from the list of reactions in the database itself; the
Pathway Tools software prepares this set of reactions through an
extensive process of excluding reactions which are unbalanced or
otherwise undesirable while expanding reactions with classes of
compounds as products or reactants into sets of possible specific
instantiations which respect conservation of mass
\cite{Latendresse2012}. Working with the Pathway Tools FBA reaction
set (rather than, e.g, an SBML export of the CornCyc database) allows
us take advantage of this pre-processing; however, it comes at the
cost of needing to reintroduce into the FBA model many reactions which
are present in the CornCyc database but are excluded from the FBA
export for one reason or another.

Reaction data was extracted from the FBA output file, and reactions were
translated to refer to species by their CornCyc frame ID (to allow
easy reference to the database and comparison with previous work, and
avoid possible ambiguities.) Reactions were then added and removed from 
the model as described below. 
 
\section{Discarding reactions}
\subsection{Polymerization reactions}
Pathway Tools attempts to include an expanded representation of
certain polymerization reactions in the exported FBA model, but this
function is considered experimental \cite{PathwayToolsManual}; these
reactions were ignored. Note that some reactions representing polymer
growth were added manually later in the process.
\subsection{ATPases}
We removed all reactions from CornCyc which have the effective stoichiometry
\begin{quote}
\texttt{
\{`ADP': 1.0, `ATP': -1.0, `PROTON': 1.0, `WATER': -1.0, `|Pi|': 1.0\}
}
\end{quote}
There are nine such reactions:
\begin{itemize}
\item \texttt{RXN-11109},
\item \texttt{3.6.4.6-RXN},
\item \texttt{RXN-11135},
\item \texttt{RXN0-1061},
\item \texttt{ADENOSINETRIPHOSPHATASE-RXN},
\item \texttt{3.6.4.4-RXN},
\item \texttt{3.6.4.9-RXN},
\item \texttt{3.6.4.5-RXN},
\item \texttt{3.6.4.3-RX}N
\end{itemize}
all treated as reversible by the Pathway Tools export
procedure. Typically these are simplified representations of the
metabolic effect of enzymes whose complete function is outside the
scope of the database, as, for example, EC 3.6.4.3, the
microtubule-severing ATPase.

\label{MaintenanceATPase} 
In their place, we added a single generic ATPase reaction to represent
cellular maintenance costs, etc., with no associated genes. 

\subsection{Reactions involving generic electron donors and acceptors}
Numerous reactions in the database are written with generic
representations of electron carrier species (`a reduced electron
acceptor', `an oxidized electron acceptor'). Most of these reactions
are outside the areas of emphasis of the model (e.g., brassinosteroid
biosynthesis), have no curated pathway assignment, or also appear in
forms which do specify the electron carrier species (e.g., the generic
nitrate reductase reaction, \texttt{NITRATEREDUCT-RXN}, vs
\texttt{NITRATE-REDUCTASE-NADH-RXN},) and so could be safely
neglected. A small set of exceptions identified in early drafts
included reactions of fatty acid synthesis, handled as discussed
below, and proline dehydrogenase, \texttt{RXN-821}, catalyzed by a
mitochondrial-membrane-bound flavoprotein which donates electrons
directly to the mitochondrial electron transport chain
\cite{Elthon1982}.  Because we have not thoroughly compartmentalized
amino acid metabolism, we implemented this reaction as donating
electrons to NAD\textsuperscript{+} instead.

\subsection{Duplicates}
A number of other reactions were removed because they appeared to be
exact (possibly unintentional) duplicates, down to gene associations,
of other reactions in the database; or because they were being
replaced by modified forms as discussed below.  These are given in
\texttt{reactions\_to\_remove.txt}.

\subsection{Non-metabolic reactions}
A number of reactions present in CornCyc were removed because the
database indicated, e.g. through the Enzyme Commision summary for the
relevant EC number, that they were primarily involved in
extrametabolic functions (e.g., cell movement, regulation).  These
included the GTPases \texttt{RXN-5462}, \texttt{3.6.5.2-RXN}, and
\texttt{3.6.5.5-RXN}.

\subsection{Glucose-6-phospate}
In the reduced model (discussed below) only one reaction,
myo-inositol-1-phosphate synthase, consumes the generic
glucose-6-phosphate species, rather than alpha-G6P or beta-G6P. To
ensure that this reaction was appropriately connected to other G6P
producing and consuming reactions we manually split it into two
instances, one for alpha-G6P and one for beta-G6P.

\subsection{UDP-glucose}
For apparently all reactions in CornCyc involving UDP-glucose, the
instantiation procedure produced one version involving generic
UDP-D-glucose and one version involving UDP-alpha-D-glucose, the only
child of the UDP-D-glucose class. UDP-alpha-D-glucose participated
in almost no reactions other than these instantiations (in the 
reduced model, described below, only one: UDP-sulfoquinovose synthase,
EC 3.13.1.1). As such there is little to distinguish the generic
and specific versions of the reactions, which add complexity to the model
and degeneracy to optimization predictions without providing significant
information about the function of the system, so we removed the
specific versions and changed the UDP-sulfoquinovose synthase to act on 
a generic UDP-D-glucose substrate.

\section{Minor revisions to achieve basic functionality}
\subsection{Mitochondrial electron transport chain} 
The CornCyc representation of the mitochondrial electron transport
pathway (\texttt{PWY-3781}, plus the mitochondrial ATPase
(\texttt{ATPSYN-RXN}, EC 3.6.3.14)) was adjusted. Some reactions
excluded from the initial Pathway Tools export because the balance
state of reactions involving cytochrome C could not be determined were
readded manually; ubiquinones/ubiquinols were uniformly represented as
ubiquinone-8/ubiquinol-8, and compartments were assigned to reactants
and products to properly represent the transport of protons between
the mitochondrial matrix and the mitochondrial intermembrane space. In
CornCyc, as in MetaCyc and other related databases, transport of
protons across the membrane is represented explicitly for complex I
but not for complex III and complex IV; in agreement with the standard
description of mitochondrial electron transport (see,
e.g., \cite{Brownleader}) proton transport was added to these reactions
with a stoichiometry of 2 H+/e- for complex III and 1 H+/e- for
complex IV.  The stoichiometry of complex IV was further adjusted to
include the H+ from the mitochondrial matrix that binds to oxygen to
form water. 

\subsection{Photosynthesis: light reactions}
Similarly, some modifications were made to the light reactions of
photosynthesis (\texttt{PWY-101}). Reactions involving plastocyanins were not
exported and were added manually; a chloroplastic ATP synthase and a
reaction describing cyclic electron transport around PS I were added;
and the stoichiometry of proton transport was adjusted in accordance
with recent literature, assuming a Q cycle and ratio of 14 H+/3 ATP
for the chloroplast ATP synthase \cite{Allen2003}.

Reduction of oxygen to superoxide at photosystem I (the Mehler
reaction) was added to allow flux through the pathways of
chloroplastic reactive oxygen species detoxification: superoxide
dismutase and the ascorbate-glutathione cycle, including a reaction
representing the direct, non-enzymatic reduction of
monodehydroascorbate by ferredoxin \cite{Asada1999,
  FoyerHarbinsonChapter}.

\subsection{Key reactions in biomass component production and nutrient uptake}
Several components of biomass required either manual adjustment of
reactions from the database or the addition of abstract synthesis
reactions summarizing the behavior of pathways which could not easily
be represented in more detail.

\subsubsection{Starch}
Starch synthase (\texttt{GLYCOGENSYN-RXN}) is not exported from CornCyc by
default (it is a polymerization reaction, and marked as unbalanced in
the PGDB); it was added manually in a form that produces the
equivalent of one 1,4-alpha-D-glucan subunit.

The starch branching enzyme EC 2.4.1.18 (\texttt{RXN-7710}) is not exported
from CornCyc by default (one reactant, starch, has an unspecified
structure); it was added manually as
\[\text{a 1,4-alpha-D-glucan subunit} \to \text{an amylopectin subunit}\]
Note that this stoichiometry is not intended to suggest that
the branching enzyme introduces branches at each subunit.

CornCyc provides a detailed reconstruction of the reactions of starch
degradation (\texttt{PWY-6724}) which is by nature difficult to convert to a
form suitable for FBA calculations, as many of the stoichiometry
coefficients are undefined. To incorporate the effects of the
glucan-water and phosophoglucan-water dikinases, for example, we would
need to specify how many glucosyl residues must be phosphorylated (and
then dephosphorylated) to produce ``an exposed unphosphorylated,
unbranched malto-oligosaccharide tail on amylopectin'' of a given
length; modeling the release of maltose from that tail would require
an estimate of the typical unbranched length of such tails,
etc. Rather than estimate average values for these parameters, we
divide the reactions of the pathway into two types: those which
condition starch for depolymerization , and actual depolymerization
reactions. The first class (the dikinases above plus isoamylase) share
the abstract stoichiometry
\[
\text{a starch subunit} \to \text{an exposed starch subunit}
\]
(neglecting any ATP costs), 
while the second class (beta amylase and disporportionating enzyme) 
convert exposed starch subunits to sugars appropriately. 

The beta-maltose releasing reactions of the starch degradation pathway
in CornCyc have no associated genes. We temporarily associated these
reactions with the beta amylase record in the database (\texttt{RXN-1827}, EC
3.2.1.2) pending further review. 

During transient starch degradation, beta-maltose and glucose are
exported into the cytosol, where maltose is split, releasing one
glucose molecule and donating one glucosyl residue to a cytosolic
heteroglycan, from which it may be released in turn as
glucose-1-phosphate \cite{Fettke2009}.  In Arabidopsis, specific
enzymes (DPE2 and PHS2) are known to be implicated in this process
\cite{Streb2012}. In simulations with this CornCyc-based FBA model we
find the typical mode of breakdown of cytosolic maltose is to
alpha-D-glucose and alpha-D-glucose-1-phosphate via
\texttt{AMYLOMALT-RXN},
\[ \text{maltotriose} + \text{beta-maltose} \to \text{maltotetraose} + \text{beta-D-glucose} \]
and \texttt{RXN0-5182}, 
\[ \text{maltotetraose} + \text{phosphate} \to \text{maltotriose} +
     \text{alpha-D-glucose-1-phosphate} \]
effectively the standard pathway but with maltotriose/maltotetraose
playing the role of the cytosolic heteroglycan pool. This
approximation leads to a reasonable effective stoichiometry but it is
possible that the genes associated with these reactions do not
accurately represent the genes involved in the true underlying
process; we have not systematically looked for maize counterparts of
the Arabidopsis genes, for example.

\subsubsection{Cellulose}
The UDP-forming cellulose synthase, EC 2.4.1.12, is not exported from
CornCyc by default (it is a polymerization reaction, and marked as
unbalanced in the PGDB); it was added manually in a form that produces
the equivalent of one subunit.
\subsubsection{Hemicellulose} 
Similarly, the following hemicellulose polymerization reactions were added manually:
\begin{itemize}
\item 1,4-beta-D-xylan synthase, EC 2.4.2.24,
\item reactions \texttt{RXN-9093} (EC 2.4.2.-) and \texttt{RXN-9094} (EC 2.4.1-),
  representing the addition of arabinose and glucuronate to xylan to
  form arabinoxylan and glucuronoxylan respectively (note that the
  corresponding subunits notionally consist of one xylan subunit plus
  arabionose/glucuronate),
\item glucomannan synthase, EC 2.4.1.32,
\item \texttt{RXN-9461} (EC 2.4.2.39), representing the addition of xylose to a
  glucan (as implemented, cellulose) to form xyloglucan (again, the
  corresponding effective subunit corresponds to one glucan subunit
  plus xylose)-- note this representation ignores the previous step in
  CornCyc's xyloglucan biosynthesis pathway, xyloglycan
  4-glucosyltransferase (EC 2.4.1.168).
\end{itemize}

In addition to these explicit descriptions of hemicellulose formation
from CornCyc, we added generic reactions representing the donation of
the following sugar residues from activated donor molecules to
unspecified generic polysaccharides: 
\begin{itemize}
\item arabinose (from UDP-L-arabinose)
\item galactose (from GDP-L-galactose)
\item galacturonate (from UDP-D-galacturonate)
\item glucose (from UDP-glucose)
\item glucuronate (from UDP-D-glucuronate)
\item mannose (from GDP-alpha-D-mannose)
\item xylose (from UDP-alpha-D-xylose)
\end{itemize}
These reactions allow the model to represent flux of these sugars
towards hemicelluloses or other polysaccharides without explicit
synthesis pathways in CornCyc, or the construction of a hemicellulose
term in the biomass equation in terms of the overall composition of
hemicellulose without reference to specific synthesis reactions, as in
our adaptation of the biomass reaction of iRS1563 (see the biomass
reaction discussion, below.)

\subsubsection{Miscellaneous cell wall components}
The following additional cell wall comoponent production reactions
from CornCyc were added manually:
\begin{itemize}
\item \texttt{2.4.1.43-RXN}, representing the formation of homogalacturonan
  from galacturonate
\item \texttt{RXN-9589} (EC 2.4.2.41), representing the addition of xylose to
  homogalacturonan to form xylogalacturonan (note the resulting
  xylogalacturonan subunit notionally consists of one galacturonate
  plus xylose)
\item \texttt{13-BETA-GLUCAN-SYNTHASE-RXN} (EC 2.4.1.12), representing the
  formation of callose from glucose.
\end{itemize}

Suberin production is not represented in CornCyc in detail but
pathways for the synthesis of three key precursors,
N-feruloyltyramine, octadecenedioate, and docosanediotate, are
provided. Sinks for N-feruloyltyramine and octadecenedioate were added
to the model to represent the flow of material towards suberin
production; docosanedioate was neglected because no genes are
associated with the reactions of its synthesis
pathway. N-feruloyltyramine may be produced from trans-caffeate via
either ferulate or caffeoyl-CoA; the branch through ferulate was
initially dropped from the reduced version of the model used for data
analysis because it relies on trans-feruloyl-CoA synthase, EC
6.2.1.34, which has no associated genes, but it was preserved in
subsequent versions of the model because high expression levels for
caffeate O-methyl\-transferase suggest this branch is indeed active.

(In CornCyc, the tyramine N-feruloyl\-transferase that produces
N-feruloyl\-tyramine from feruloyl-CoA could also catalyze the
production of other hydroxy\-cinnamic acid tyramine amides
(cinnamoyl\-tyramide, sinapoyl\-tyramide, p-coumaroyl-tyramine) but we have
neglected these for now.)

\subsubsection{Fatty acids and lipids}
Plant fatty acid and lipid biosynthesis is rich in complexity
(see, e.g., \cite{Li-Beisson2013}), and attempting to describe it in
the FBA model at the level of detail at which it is currently
understood would require a daunting number of reactions among the
species representing the combinations of lipid head groups and acyl
chains.  Though CornCyc presents some pathways of lipid metabolism at
such a high resolution, we have adopted a simplified approach which
aims to include enough detail to allow the model to:
\begin{itemize}
\item predict based on RNA-seq data the total flow of biomass
  into fatty acids and lipids
\item coarsely predict differences in the types of lipids and fatty
  acids produced, based on RNA-seq data
\item approximately preserve the iRS1563 biomass equation.
\end{itemize}
The model describes in detail the sequence of reactions by which fatty
acids up to lengths of 16 and 18 are synthesized in the chloroplast
(though currently these reactions occur in the cytoplasmic
compartment!), and the formation of oleate (as oleoyl-ACP) by the
stearoyl-ACP desaturase (\texttt{PWY-5156}; \cite{Ohlrogge1995,
  Li-Beisson2013}). In practice, these fatty acids may then enter the
`prokaryotic' pathway of glycerolipid synthesis in the chloroplast or
leave the chloroplast and enter the `eukaryotic' pathway of
glycerolipid synthesis in the endoplasmic reticulum, with further
desaturation of the acyl chains occurring after their incorporation
into lipids. 

We simplify this process by effectively decoupling the synthesis of
different types of lipids (as distinguished by head groups) from the
desaturation of their associated acyl chains. Reactions from lipid
synthesis pathways are implemented as if all lipid species had one
16:0 and one 18:1 acyl chain, by implementing the glycerol-3-phosphate
O-acyltransferase and 1-acylglycerol-3-phosphate O-acyltransferase
reactions (RXN-10462 and 1-ACYLGLYCEROL-3-P-ACYLTRANSFER-RXN), written
in the database with generic acyl-acp substrates, with oleoyl-ACP and
palmitoyl-ACP as substrates respectively. (This corresponds to the
prokaryotic pathway; in the eukaryotic pathway oleoyl-CoA and
palmitoyl-CoA would supply the acyl groups for diacylglycerol
formation instead \cite{Li-Beisson2013}. However the same genes are
associated with the reactions of diacylglycerol synthesis in the two
pathways (\texttt{PWY-5667}; \texttt{PWY0-1319}) in CornCyc and so
they cannot be distinguished based on expression data alone; we have
chosen one arbitrarily.)

This supply of diacylglycerol is sufficient to allow, without further
modification to the CornCyc FBA export, the synthesis of a variety of
lipids, including:
\begin{itemize}
\item phosphatidylcholine, phosphatidylethanolamine,
  phosphatidylglycerol, phosphatidylinositol;
\item sulfoquinovosyldiacylglycerol.
\end{itemize}
UDP-glucose epimerase is exported from CornCyc in the
UDP-glucose-producing direction by default; we allowed it to run in the
reverse direction as well, consistent with literature evidence
\cite{Dormann1998,Brenda5.1.3.2}, which allowed the production of
mono- and digalactosyldiacylglycerol.

In sphingolipid metabolism, dihydrosphingosine, 4-hydroxysphinganine
and sphinganine 1-phosphate may be produced, and sink reactions were
added for them.  Production of the ceramides and their derivatives
would require the choice of a particular fatty acid source for the
sphinganine acyltransferase, written by default with the generic
substrate `a long-chain acyl-coA'; per the CornCyc description page
for \texttt{PWY-5129}, in leaf sphingolipids C20 to C26 fatty acids are
typical. Currently, the FBA model lacks a detailed implementation of
production of very long chain fatty acids by elongation (a generic
representation is present in CornCyc), so no supply of C20-26 fatty
acids is available. We have deferred this issue to future work.

Separately, we model the desaturation of oleate to linoleate 
and linolenate and palmitate to palmitoleate. These (along with 
palmitate and stearate) are the fatty acid components
of the iRS1563 biomass reaction, which originally incorporated
them as triglycerides; our modified biomass equation consumes
free fatty acids, rather than attempt to specify the precise
ratios in which they are to be found in different lipid species
in the leaf. 

The CornCyc pathways for linoleate and linolenate produce them as
lipid linoleoyl groups and lipid linolenoyl groups respectively,
incorporated in generic lipid molecules; to allow these reactions to
balance, and to provide linoleate and linolenate for the biomass
reaction, we added lipases which release free linoleate/linolenate
from the lipid linoleoyl and lipid linolenoyl groups, regenerating the
pool of generic `lipid' species (which participate only in the
linoleate pathway, within the FBA model.)  Note, however, that other
reactions within the model but outside the indicated synthesis
pathways are capable of producing linoleate and linolenate as well.

CornCyc includes no complete pathway for the production of palmitoleic
acid; as there is experimental evidence it is produced in maize leaves
(see the discussion of the biomass equation, below) we introduced the
acyl-ACP $\Delta$9-desaturase reaction from the palmitoleate
biosynthesis pathway of AraCyc (\texttt{RXN-8389}, 1.14.99.-), producing
palmitoleoyl-ACP from palmitoyl-ACP \cite{AraCycPWY-5366}, which
restores this functionality (in combination with the palmitoleoyl-ACP
hydrolase, \texttt{RXN-9550}, which is present in CornCyc.) Note that there is
some evidence that the stearoyl-ACP desaturase enzyme may also
catalyze this reaction \cite{Gibson1993}.

The oleoyl-acyl carrier protein hydrolase (EC 3.1.2.14) from CornCyc
is unbalanced with respect to hydrogen; a version with an additional
proton on the right hand side was added manually.

The $\Delta$9-desaturase and the desaturases producing linoleate and
linolenate (\texttt{RXN-9667} and \texttt{RXN-9669}) were written
originally with generic electron donor and acceptor species.  Initial
review of the extensive literature on plant fatty acid desaturation
suggests that the electron source for desaturases depends on their
location within the cell, with chloroplastic desaturases accepting
electrons from ferredoxin while desaturases in the endoplasmic
reticulum accept electrons from NADH via cytochrome b5 or fused
cytochrome domains (see, eg, \cite{Sperling1995, Harwood1996,
  Shanklin1998}.) As discriminating between chloroplastic and
extrachloroplastic fatty acid desaturation is not a high priority for
the model, NADH was used as the sole electron donor for all three of
these reactions.

The ferredoxin-dependent stearoyl-ACP desaturase \texttt{RXN-7903},
not exported from the database by default because it is marked as
unbalanced, was added in a form adjusted for hydrogen and charge
balance. Ferredoxin-NADP oxidoreductase was made reversible to ensure
NADPH can drive this reaction in the dark, as is observed
\cite{Shanklin1998}.

\subsubsection{Nucleic acids polymerization}
Reactions representing the pyrophosphate-releasing incorporation of
(d)NTPs into RNA and DNA were added and associated with the
DNA-directed DNA polymerase and DNA-directed RNA polymerase reactions
in the database. (In each case, it is assumed that all nucleotides
occur with equal frequency.)

\subsection{Ascorbate-glutathione cycle}
To allow the NADPH-monodehydroascorbate reductase reaction to
function in the cycle as curated, we split the L-ascorbate peroxidase reaction
(EC 1.11.1.11) into its two subreactions, which by default are not exported
in the FBA problem. 

\subsection{Gamma-glutamyl cycle}
The gamma-glutamyltransferase was lumped together with 
\texttt{GAMMA-GLUTAMYL\-CYCLO\-TRANSFERASE-RXN}, originally written in terms
of the instanceless class `\texttt{L-2-AMINO-ACID}' which appeared in no
other stoichiometries in the FBA export, and the dipeptidase \texttt{RXN-6622},
which is the only reaction that can consume the cysteinylglycine product of the
gamma-glutamyltransferase, forming a combined reaction which can carry flux.
The combined reaction retained the gene associations of the gamma-glutamyltransferase,
as the other two reactions have no associated genes.

\subsection{Methionine synthesis from homocysteine}
The methionine synthase reaction of CornCyc's methionine biosynthesis
pathway, \texttt{HOMOCYSMET-RXN}, EC 2.1.1.14, specifically requires
5-methyl\-tetra\-hydro\-pteryl\-tri-L-glutamate as a
cofactor. Polyglutamylation of folates is present in CornCyc in an
abstract representation (with tetra\-hydro\-folate synthase catalyzing the
addition of a glutamyl group to a 5-methyl\-tetra\-hydro\-pteryl with $n$
glutamyl groups); we have not converted this into an explicit
representation in the FBA model. Instead, \texttt{HOMOCYSMETB12-RXN},
EC 2.1.1.13, acts to produce methionine from homocysteine; the effects of 
this possible inaccuracy on the behavior of the rest of the network should be 
limited.

\subsection{Basic import and export}
The following species are given overall import/export reactions:
\begin{itemize}
\item \texttt{WATER}
\item \texttt{CARBON-DIOXIDE}
\item \texttt{OXYGEN-MOLECULE}
\item \texttt{PROTON}
\item \texttt{NITRATE}
\item \texttt{SULFATE}
\item \texttt{|Pi|}
\item \texttt{|Light|}
\item \texttt{MG+2}
\end{itemize}

These reactions exchange species inside the cell with species in meaningfully
labeled compartments where possible (eg, oxygen and CO\textsubscript{2} are exchanged
with the intercellular air space, mineral nutrients with the xylem, etc.)

In addition, to facilitate exchange among compartments in the
whole-leaf model, a number of exchanges with a phloem compartment were
set up: these included sucrose, glycine (as a representative of the
amino acids detected in maize phloem sap by Ohshima et al
\cite{Ohshima1990},) and the potential phloem sulfur transport
compound glutathione \cite{Bourgis1999}.

Note that these reactions should be inactive, or restricted to the exporting
direction only, when not modeling transport within the leaf (except for sucrose,
where a free supply should be allowed in heterotrophic conditions.)

\subsection{Defining the biomass components}
Two types of biomass reactions are added to the model: \begin{itemize}
\item Sinks for individual species, for simulations (e.g, fits to
  RNAseq data) where the relative rates of production of different
  components are unknown. The species given such sinks are listed in
  \texttt{biomass\_components.txt}. 
\item A set of reactions producing a combined biomass species, made up
  of assorted components in fixed proportions, for simulations where
  the maximum rate of production of biomass is of interest, and an
  approximately realistic biomass composition needs to be enforced
  directly. These reactions were taken with minor modifications from
  \cite{Saha2011}; their adaptation is described below and they are
  listed in are listed in \texttt{adapted\_irs1563\_biomass.txt}.
 \end{itemize}
To conceptually and practically separate these types of biomass reactions,
which in general should not both be active in any one calculation, the
biomass species they produce are located within two separate abstract
biomass compartments in the SBML model.

In general, the biomass sink reactions have no gene associations, but
an exception was made for the twenty reactions representing incorporation
of amino acids into protein, which inherit the gene associations of the
corresponding tRNA ligase reactions in CornCyc. (In principle these
could be distinguished from sink reactions representing the expansion of
free amino acid pools as cells grow and divide, but we have ignored this
issue for now.)

Note that, to support the adapted iRS1563 biomass equation, a reaction
representing the production of free galactose from GDP-L-galactose was
introduced (otherwise, release of galactose from UDP-galactose was
catalyzed by two reactions in the pathways of indole-3-acetyl-ester
conjugate biosynthesis and indole-3-acetate activation, likely not a
major route for carbohydrate production.) Free galactose is not included
in the individual biomass species used for data fitting.

\section{Compartmentalization}
Approaches differ to the subcellular compartmentalization in FBA
models of eukaryotes, ranging from the assignment of compartments to a
few key pathways known to function primarily outside the cytosol, as
in the mitochondrial and chloroplastic ``modules'' of AraMeta
\cite{Poolman2009} and RiceMeta \cite{Poolman2013} to the extremely
comprehensive, data-driven approach of \cite{Mintz-Oron2012}.  Here,
we did not attempt to comprehensively assign reactions to their proper
compartments; instead, we started with a modular approach similar to
\cite{Poolman2013} in which some core metabolic pathways were
compartmentalized (in our case, the TCA cycle and mitochondrial
electron transport chain in the mitochondrion, the light reactions of
photosynthesis, Calvin cycle, and some reactions of the C4 and
photorespiratory pathways in the chlorophyll, and some reactions of
the photorespiratory pathway in the peroxisome, with transport
reactions added as necessary.)

We then refined the compartment assignments of other reactions and
pathways as needed to permit key metabolic functions and
compartmentalize a limited number of additional reactions whose
incorrect assignment to the cytosol we judged particularly likely to
lead to misleading results.

More details on individual compartmentalization choices and transport
reactions are given below.

\subsection{Intracellular transport}
Sources (beyond those detailed below) informing the addition of
intracellular transport reactions in the model included the transport
reactions present in AraMeta \cite{Poolman2009}, reviews of
photorespiratory metabolism with attention to compartmentalization
\cite{Reumann2006,Foyer2009}, a review of chloroplast transporters
\cite{Weber2011}, and a review of transport processes in C4
photosynthesis \cite{Brautigam2011}.

In most cases we have not tried to reflect the mechanisms of the
transport systems, where those are known, in any detail (exceptions
include the triose phosphate-phosphate and PEP-phosphate transporters
across the chloroplast envelope), nor have we associated genes with
the transporters, even when they are known. Future work should pay
greater attention to this aspect of the system.

\subsection{Photorespiratory pathway}
Following \cite{DouceHeldt} we assumed that reducing power was
supplied to the peroxisome through an oxaloacetate-malate shuttle and
NAD(H)-dependent malate dehydrogenase, and added an
oxaloacetate-malate antiporter and a copy of \texttt{MALATE-DEH-RXN}
to the peroxisome. Reactions of the pathway were localized following
\cite{Reumann2006} and \cite{Foyer2009}. Note that glycine
decarboxylase was assigned exclusively to the mitochondrion, while
serine hydroxymethyltransferase was present in both the mitochondrion
and the cytoplasm, where it plays a role in one-carbon metabolism
\cite{Hanson2001}.

\subsection{Various ferredoxin-consuming pathways}
The model includes several pathways or reactions (e.g., sulfite and
nitrite reduction and the chlorophyll cycle) which rely on ferredoxins
for reducing power, and are localized to the chloroplast, where, in
the light, reduced ferredoxins may be supplied by the photosynthetic
electron transport chain.

Rather than assign the reactions of these pathways to compartments
appropriately, we added a reaction exchanging reduced ferredoxins and
oxidized ferredoxins across the chloroplast boundary to supply
ferredoxin-driven pathways in the cytosol. We emphasize that this is a
convenient simplification and is not intended to represent a realistic
mechanism.

\subsection{Ascorbate production}
The L-galactonolactone dehydrogenase responsible for the final step of
the ascorbate production pathway in CornCyc reduces cytochrome C and
has been experimentally localized to the mitochondrial inner membrane,
with its catalytic site facing outwards, into the intermembrane space
\cite{Bartoli2000}.  As the outer membrane is generally permeable to
small molecules we have treated this reaction as acting directly on
cytoplasmic galactonolactone and ascorbate. A sink for ascorbate as a
biomass component was added, as it is found in substantial quantities
in leaves (see, e.g., \cite{Foyer1983,Smirnoff1996}.)

\subsection{Ascorbate-glutathione cycle}
This cycle is present in multiple cellular compartments; in the model
we included only cytosolic and chloroplastic instances (of which only
the chloroplastic was ultimately expected to be relevant, as there was
no supply of superoxides in the cytosol.)  Note that none of the genes
associated with monodehydroascorbate reductase could be assigned to
the chloroplast under the rules described below: two had curated
location in the peroxisome while \texttt{GRMZM2G320307} had no curated
location and TargetP prediction of mitochondrial
(\texttt{GRMZM2G320307\_P01}) and cytoplasmic
(\texttt{GRMZM2G320307\_P02}, \texttt{GRMZM2G320307\_P03})
locations. Reduction of monodehydroascorbate may also proceed
non-enzymatically (see above) so this (enzymatic) reaction was removed
from the chloroplast in favor of direct reduction by ferredoxin.

\section{Gene associations for compartmentalized reactions}
Where a reaction was present in more than one compartment-- that is,
when two or more reactions in different compartments were associated
with the same reaction record in CornCyc-- we examined the genes
associated with those reactions in CornCyc and assigned them to the
instance of the reaction in the most appropriate compartment, as far
as possible.

Where the Plant Proteome Database \cite{Sun2009} provided manually
curated location assignments for genes, those were used; otherwise, we
used automatic location predictions by TargetP \cite{Emanuelsson2000}
or in some cases referred to the gene's annotation (both also provided
by PPDB.) In general we assumed the appropriate location for a gene
product was the cytoplasmic compartment absent a specific prediction
of localization in the chloroplast, mitochondrion, or
peroxisome. Where proteins were predicted to occur in a compartment
where an no instance of a particular reaction was present, those gene
associations were generally dropped from the model.

When a gene was associated with a reaction in more than one
compartment and also a reaction present in only one compartment, in
general the association with the reaction in only one compartment was
dropped, except for reactions which we believed based on literature
evidence (including comments in CornCyc and PPDB) were assigned to the
cytoplasmic compartment only because our compartmentalization process
was incomplete.

Some details on the judgment calls made in this process are provided in the 
comments to the file \texttt{gra\_overrides.txt}; we comment here on a few 
unusual cases.

\subsection{NADH dehydrogenases}
Cyclic electron transport around Photosystem I may occur through the
chloroplast NADH dehydrogenase complex or an alternate pathway which
in Arabidopsis involves PGR5 \cite{Munekage2004, Shikanai2007}.  In C3 plants
the PGR5-dependent pathway may play the major role in tuning the
photosynthetic ATP/NADPH ratio, while the NADH dehydrogenase pathway
is implicated in stress responses \cite{Shikanai2007}.  In contrast, in C4
plants the expression of the chloroplast NADH-dehydrogenase appears to
correlate with photosynthetic ATP demand, while PGR5 expression does
not, suggesting it is the NADH-dehydrogenase CET pathway which allows
increased the increased ATP production required by the C4 system
\cite{Takabayashi2005}. Thus, genes associated in CornCyc with 
the NADH dehydrogenase reaction for which a chloroplast location was
predicted were reassociated with the model's cyclic electron transport 
reaction (despite the fact that our somewhat abstract cyclic
electron transport reaction may not accurately represent the
biochemistry of the NADH-dependent pathway.) 

\subsection{Pyruvate dehydrogenases}
In practice, pyruvate dehydrogenase complexes are found in the
mitochondrion and chloroplast, but here we have not fully
compartmentalized the chloroplastic pyruvate dehydrogenase and the
pathways it supplies, instead leaving it in the cytosol. Thus, genes
associated with the reactions of the complex with predicted
chloroplast localization were associated instead with the cytosolic
version. Genes with no curated or predicted location were left
associated with both forms (splitting their expression data between
them, in the fitting process.)

\section{Testing and consistency checking}
The compartmentalized single-cell model was checked in detail for
conservation violations by testing the feasibility of net production
or consumption of a unit of each internal species with all external
transport and biomass sink reactions suppressed.

Where such production was found feasible, the reactions involved were
carefully inspected and stoichiometry coefficients adjusted to restore
balance if necessary. In practice, this led only to the correction of
erroneous reactions added by hand; 
as expected, no balance issues were found with reactions exported from
CornCyc.

In the final version, 
no such unrealistic processes are possible in the model under normal
conditions. (Note that the species representing light input may be
consumed in isolation, but the use of light energy to drive a futile
cycle is not unrealistic, though we have not examined the details of
the process found by the consistency checker in any detail.)  Of
course, demonstrating that no such production/consumption is feasible
does not guarantee that all reactions in the model are properly
balanced.

Testing also verified that all individual biomass sink reactions, and the 
combined biomass reaction, could proceed at nonzero rates.

\section{SBML export}
\label{sbml}
\subsection{Component names}
SBML distinguishes a component's name from its ID.  Reactions and
species in the SBML model were given name attributes according to the
by calling the Pathway Tools \texttt{get\_name\_string} function on
the frames in the database from which they derive, if any. The IDs of
the SBML components were derived from the frame handles, replacing
special characters with underscores as necessary to conform to the
SBML sID standard.

Note that for some reactions in CornCyc, the result of
\texttt{get\_name\_string} is an EC number different from the EC
number indicated by the label of the frame (e.g,
\texttt{2.7.1.133-RXN}, for which `EC 2.7.1.159' is returned.) The
frame in CornCyc (if any) from which each reaction in the SBML model
is ultimately derived is preserved as a comment in the reaction's
Notes element, to resolve any ambiguity.

\subsection{Gene annotations}
Each reaction in the FBA model associated with a particular parent
frame in CornCyc was given an association rule that combined all genes
associated with that reaction in CornCyc, as well as all genes
associated with all generic reactions of which the parent reaction is
a specific form, in a logical `or' relationship, stored in the
reaction's Notes element per the COBRA standard.

\section{Model refinement}
\label{refinement}
\subsection{Phosphoribulokinase}
In early attempts to fit the model to the leaf gradient data, high
costs were associated with the mesophyll phosphoribulokinase reaction
in the source tissue when the bundle sheath CO\textsubscript{2} level was
high. We noted that in CornCyc 4.0 several genes were associated with
both PRK and glyceraldehyde-3-phosphate dehydrogenase. To clarify the
role of these genes we referred to annotations in the Plant Proteome
Database \cite{Sun2009} and best hits in the Conserved Domain
Database (\cite{Marchler-Bauer2004}, accessed through NCBI.) Of the
eight genes associated with PRK in CornCyc, three
(\texttt{GRMZM2G039723}, \texttt{GRMZM2G337113},
\texttt{GRMZM2G162845}) appeared to encode GAPDH enzymes (per PPDB
annotations and the presence of \texttt{Gp\_dh\_N} and
\texttt{Gp\_dh\_C} domains), three (\texttt{GRMZM2G162529,
  GRMZM2G463280, GRMZM2G026024}) appeared encode to encode genuine
phosphoribulokinases (per PPDB annotations and the presence of PRK
domains), and two appeared to encode CP12-type regulatory proteins,
with no obvious evidence for any individual protein sharing more than
one of these roles. The regulatory role of CP12 does involve forming a
complex with PRK and GAPDH, but this reduces, rather than enhancing or
enabling, their individual activities \cite{Lopez-Calcagno2014}. We
removed the PRK associations of the GAPDH and CP12 genes from our
model. PPDB assigned these three GAPDH genes to a plastidic location
based on experimental evidence, so we associated them with those
reactions exclusively (removing associations with the cytosolic
instances of EC 1.2.1.13 and/or EC 1.2.1.12.)

\section{Biomass equation}
\label{biomass}
We developed a biomass equation following that used in
\cite{Saha2011}.  Our calculations are based on supplementary file S4
of that paper\footnote{Specifically,
  \texttt{journal.pone.0021784.s004.xls}, as downloaded from the PLoS
  One web site 20 November 2013}, in particular sheet 2,
`Biomass\_rxn'.

That sheet derives a biomass equation corresponding to the production
of one gram of plant dry weight, based on literature data on biomass
composition; the description is divided into subreactions forming
(e.g.) `nitrogenous compounds', `lignin', etc., which then participate
in an overall biomass reaction.) The units of the stoichiometric
coefficients are mmol.

We have adopted most of the biomass composition assumptions of Saha et
al wholesale, with gratitude for their efforts in compiling this data
from the literature. However, we have made some minor adjustments,
resulting in a different overall stoichiometry for biomass production. 

\subsection{Fatty acids}
Saha et al represent the total lipid/fatty acid contribution to
biomass as a pool of triglycerides in proportions apparently based on
a maize oil measurement and thus probably reflective of seed
triglyceride composition.

We substitute measurements of the fatty acid content of mature maize
leaf membrane lipids \cite{Rizov2000} and write a biomass
sub-reaction which consumes the relevant free fatty acids (rather than
their derivatives in the form of triacylglycerols, membrane lipids,
etc.,) as shown in Table \ref{fatty_acid}.
\begin{table}[h]
\centering
\begin{tabular}{l|l|r|r}
Fatty acid & CornCyc compound & mol. wt. (g/mol) & mole fraction \\ \hline
palmitic & \tt{PALMITATE} & 255.42 & 0.104 \\
palmitoleic & \tt{CPD-9245} & 253.4 & 0.056 \\
stearic & \tt{STEARIC\_ACID} & 283.47 & 0.011 \\
oleic & \tt{OLEATE\_CPD} & 281.46 & 0.044 \\
linoleic & \tt{LINOLEIC\_ACID} & 279.44 & 0.132 \\
linolenic &  \tt{LINOLENIC\_ACID} & 277.43 & 0.646 \\
\end{tabular}
\caption{Fatty acid proportions in biomass.} 
\label{fatty_acid}
\end{table}
\\
Weighting the molecular weights by the mole fractions, we find one mole of 
fatty acid in appropriate proportions weighs
272.4 g. Dividing the mole fractions by the overall molar weight and
multiplying coefficients by 1000 to convert to millimoles, we arrive at the final
equation:
\begin{quote}
0.382 \texttt{PALMITATE} + 0.206 \texttt{CPD-9245} + 0.04 \texttt{STEARIC\_ACID} + 
0.162 \texttt{OLEATE\_CPD} + 0.485 \texttt{LINOLEIC\_ACID} + 2.372 \texttt{LINOLENIC\_ACID} = 
\tt{fatty\_acids\_biomass}
\end{quote}
where the left-hand side represents 1 g. 

Fractions add to less than 1.0 because we ignore trace (mole fraction
$\leq$ 0.01) amounts of C14:0 and C20:0 fatty acids.  Note that the leaf
fatty acid composition is known to change along the developmental
gradient, so specifying any single composition is an approximation;
see \cite{Leech1973}.

\subsection{Hemicellulose}
We adopted the hemicellulose production reaction as is, using the
species added to the model for this purpose, 
`\texttt{polysaccharide\_[sugar]\_unit}'.
The resulting equation is:
\begin{quote}
0.548 \texttt{polysaccharide\_arabinose\_unit} + \\ 
1.248 \texttt{polysaccharide\_xylose\_unit } + \\
0.301 \texttt{polysaccharide\_mannose\_unit } + \\ 
0.144 \texttt{polysaccharide\_galactose\_unit} + \\
3.254 \texttt{polysaccharide\_glucose\_unit} + \\
0.166 \texttt{polysaccharide\_galacturonate\_unit} + \\
0.166 \texttt{polysaccharide\_glucuronate\_unit} = \texttt{hemicellulose\_biomass}.
\end{quote}


\subsection{Total carbohydrates}
We recalculated the stoichiometries of the carbohydrate-producing
reaction to account for the differing molecular weight of our
representation of cellulose (`\texttt{CELLULOSE\_monomer\_equivalent}',
effectively a glucose molecule), account for the fact that one unit
of hemicellulose represents one gram, not one (milli)mole, and express
pectin in terms of \texttt{polysaccharide\_galacturonate\-\_unit}, reflecting a 
belief that UDP is released in the formation of pectin from UDP-D-galacturonate,
rather than retained in the polymer \cite{CornCycPWY-1061}. 

It is not clear what form the `mannose' referred to by Penning de Vries et al
should be assumed to take, as free mannose is not found in plants under
most circumstances
(see, e.g., \cite{Herold1977,Schnarrenberger1990,PlantCycMANNCAT-PWY}.)
Here we somewhat arbitrarily choose mannose-6-phopshate.
\begin{table}[h]
\centering
{\tiny
\begin{tabular}{l|l|r|c|c}
Component & Species in model & unit wt (mg) & wt fraction & units/g product\\
\hline
Ribose & \tt{RIBOSE} & 150.053 & 0.010 & 0.067\\
Glucose & \tt{GLC} & 180.063 & 0.050 & 0.278\\
Fructose & \tt{FRU} & 180.063 & 0.020 & 0.111\\
Mannose & \tt{MANNOSE-6P} & 258.120 & 0.010 & 0.039\\
Galactose & \tt{GALACTOSE} & 180.063 & 0.010 & 0.056\\
Sucrose & \tt{SUCROS} & 342.116 & 0.050 & 0.146\\
Cellulose & \tt{CELLULOSE\_monomer\_equivalent} & 180.160 & 0.400 & 2.220\\
Hemicellulose & \tt{hemicellulose\_biomass} & 1000.000 & 0.400 & 0.400\\
Pectin & \tt{polysaccharide\_galacturonate\_unit} & 193.130 & 0.050 & 0.259\\
\end{tabular}
}
\caption{Carbohydrate species in biomass.}\label{carbohydrates}
\end{table}

Table \ref{carbohydrates} shows the calculation, resulting in the equation:
\begin{quote}
0.067 \texttt{RIBOSE} + 0.278 \texttt{GLC} + 0.111 \texttt{FRU} + 0.039 \texttt{MANNOSE-6P} + 0.056 \texttt{GALACTOSE} + 0.146 \texttt{SUCROSE} + 2.220 \texttt{CELLULOSE\_monomer\_equivalent} + 0.400 \texttt{hemicellulose\_biomass} + 0.259 \texttt{polysaccharide\_galacturonate\_unit} = \texttt{carbohydrates\_biomass}.
\end{quote}

\subsection{Organic acids}
We adopt this reaction as is. In the terminology of our model, the resulting equation is:
\begin{quote}
0.556 \texttt{OXALATE} + 0.676 \texttt{GLYOX} + 1.515 \texttt{OXALACETIC\_ACID} + 0.746 \texttt{MAL} + 1.562 \texttt{CIT} + 1.724 \texttt{CIS-ACONITATE} = \texttt{organic\_acids\_biomass}.
\end{quote}

\subsection{Protein and free amino acids}
We adopt these reactions as is.
In the terminology of our model, the resulting equations are:
\begin{quote}
1.15 \texttt{L-ALPHA-ALANINE} + 0.0959 \texttt{ARG} + 0.414 \texttt{L-ASPARTATE} + 0.0313 \texttt{CYS} + 1.53 \texttt{GLT} + 0.0445 \texttt{GLY} + 0.0915 \texttt{HIS} + 0.465 \texttt{ILE} + 1.51 \texttt{LEU} + 5.71e-05 \texttt{LYS} + 0.123 \texttt{MET} + 0.314 \texttt{PHE} + 0.762 \texttt{PRO} + 0.612 \texttt{SER} + 0.175 \texttt{THR} + 0.00409 \texttt{TRP} + 0.244 \texttt{TYR} + 0.25 \texttt{VAL} = \texttt{protein\_biomass}
\end{quote}
and
\begin{quote}
0.624 \texttt{L-ALPHA-ALANINE} + 0.319 \texttt{ARG} + 0.418 \texttt{L-ASPARTATE} + 0.231 \texttt{CYS} + 0.378 \texttt{GLT} + 0.740 \texttt{GLY} + 0.358 \texttt{HIS} + 0.424 \texttt{ILE} + 0.424 \texttt{LEU} + 0.380 \texttt{LYS} + 0.373 \texttt{MET} + 0.337 \texttt{PHE} + 0.483 \texttt{PRO} + 0.529 \texttt{SER} + 0.467 \texttt{THR} + 0.272 \texttt{TRP} + 0.307 \texttt{TYR} + 0.475 \texttt{VAL} = \texttt{free\_aa\_biomass}.
\end{quote}

\subsection{Lignin}
We adopt this reaction as is. In the terminology of our model, the resulting
equation is:
\begin{quote}
2.221 \texttt{COUMARYL-ALCOHOL} + 1.851 \texttt{CONIFERYL-ALCOHOL} + 1.587 \texttt{SINAPYL-ALCOHOL} = \texttt{lignin\_biomass}.
\end{quote}

\subsection{Nucleic acids}
We adopt this reaction as is (though note that, as discussed above,
nucleotide triphosphates are not necessarily the appropriate best
representation for polymerized nucleic acids). In the terminology of
our model, the resulting equation is:
\begin{quote}
0.247 \texttt{ATP} + 0.239 \texttt{GTP} + 0.259 \texttt{CTP} + 0.258 \texttt{UTP} + 0.255 \texttt{DATP} + 0.247 \texttt{DGTP} + 0.268 \texttt{DCTP} + 0.259 \texttt{TTP} = \texttt{nucleic\_acids\_biomass}.
\end{quote}

\subsection{Nitrogenous compounds}
We use the same nitrogenous compound weight fraction breakdown, but recalculate the
stoichiometric coefficients accounting for the fact that the protein biomass,
free amino acid biomass, and nucleotide biomass species each represent one gram,
so that the appropriate stoichiometric coefficients of those species
for the production of one total gram of nitrogenous compounds are simply the
weight fractions; see Table \ref{nitrogenous}.
\begin{table}[h]
\centering
\begin{tabular}{l|l|r|c|c}
Component & Species in model & unit wt (mg) & wt fraction & units/g product\\
\hline
Amino acids & \tt{free\_aa\_biomass} & 1000.000 & 0.100 & 0.100\\
Proteins & \tt{protein\_biomass} & 1000.000 & 0.870 & 0.870\\
Nucleic acids & \tt{nucleic\_acids\_biomass} & 1000.000 & 0.030 & 0.030\\
\end{tabular}
\caption{Nitrogenous biomass breakdown.}
\label{nitrogenous}
\end{table}

The resulting equation is
\begin{quote}
0.100 \texttt{free\_aa\_biomass} + 0.870 \texttt{protein\_biomass} + 0.030 \texttt{nucleic\_acids\_biomass} = \texttt{nitrogenous\_biomass}.
\end{quote}

\subsection{Inorganic materials}
We ignore these entirely, as they play no other role in the model. (Note that 
even in iRS1563 the two species involved, potassium and chloride, participate
only in source and sink reactions.) 

\subsection{Total biomass reaction}
We drop the inorganic materials term (note that weight fractions now add to 0.95)
and recalculate the stoichiometric coefficients, accounting for the fact that
the component biomass subspecies each represent one gram; see Table \ref{total}. 

\begin{table}[h]
\tiny{
\begin{tabular}{l|l|r|c|c}
Component & Species in model & unit wt (mg) & wt fraction & units/g product\\
\hline
Nitrogenous compounds & \tt{nitrogenous\_biomass} & 1000.000 & 0.230 & 0.230\\
Carbohydrates & \tt{carbohydrates\_biomass} & 1000.000 & 0.565 & 0.565\\
Lipids & \tt{fatty\_acids\_biomass} & 1000.000 & 0.025 & 0.025\\
Lignin & \tt{lignin\_biomass} & 1000.000 & 0.080 & 0.080\\
Organic acids & \tt{organic\_acids\_biomass} & 1000.000 & 0.050 & 0.050\\
\end{tabular}
}
\caption{Breakdown of total biomass.}
\label{total}
\end{table}

The final equation is
\begin{quote}
0.230 \texttt{nitrogenous\_biomass} + 0.565 \texttt{carbohydrates\_biomass} + 0.025 \texttt{fatty\_acids\_biomass} + 0.080 \texttt{lignin\_biomass} + 0.050 \texttt{organic\_acids\_biomass} = \texttt{total\_biomass}.
\end{quote}

Saha et al additionally incorporate an ATP cost in their overall
biomass reaction, based on that used in an earlier Arabidopsis model
(AraGEM \cite{GomesdeOliveiraDalMolin2010})
 Combining this ATP hydrolysis with a sink of total biomass, we
arrive at the overall equation for biomass production and growth
(\texttt{CombinedBiomassReaction}):
\begin{quote}
1.0 \texttt{total\_biomass} + 30.0 \texttt{ATP} 
+ 30.0 \texttt{WATER} = 30.0 \texttt{ADP} + 30.0 \texttt{Pi} + 30.0 \texttt{PROTON}
\end{quote}

\subsection{Protonation}
Throughout, note that the molecular weights of species in our model
may differ somewhat from those used in the iRS1563 table because of
differing assumptions about protonation. The practical consequences
of this difference should be limited.

\subsection{Oxalate}
Early drafts of the model could not produce oxalate.  CornCyc
indicates its production as resulting only from ascorbic acid
catabolism with concomitant production of L-threonate. Recent
reviews suggest this is the primary pathway of oxalate production in
plant species which form calcium oxalate crystals, with the threonate
ultimately being oxidized to tartrate \cite{Franceschi95,
  Franceschi2005, Debolt2007}, though the pathways of production of soluble
oxalate are less clear \cite{Franceschi2005}. We found little immediate
evidence that tartrate (or threonate) is formed in maize leaves at
levels comparable to that of oxalate, or of pathways which could
further metabolize the tartrate.

Of the three reactions in iRS1563 which could produce oxalate, only
one has an associated gene: oxalate carboxylase (oxalate = formate +
CO2); KEGG R00522 (EC4.1.1.2). The gene, `ACG37538', may correspond to
\texttt{GRMZM2G103512}, whose best Arabidopsis hit is
\texttt{AT1G09560.1} (germin-like protein 5); it may thus be more
likely to be an oxalate-consuming oxidase \cite{Lane1993} than an
oxalate carboxylase, though no function was computationally predicted
for \texttt{GRMZM2G103512} in CornCyc.

We decided the available information did not allow us to accurately
model oxalate production in maize. However, to retain the iRS1563
biomass equation and ensure that mass and elemental balance was
preserved, we allowed production of oxalate from oxaloacetate by
oxaloacetase (EC 3.7.1.1; PlantCyc \texttt{OXALOACETASE-RXN},
\cite{PlantCycOXALOACETASE-RXN}). This simple reaction has been observed in fungi
\cite{Hayaishi1956} but is considered unlikely to be wide\-spread in
plants \cite{Franceschi2005}.

\section{Plasmodesmatal transport reactions}
\label{plasmodesmata} 

Species allowed to be exchanged between cell types through the
plasmodesmata included:
\begin{itemize}
\item carbon dioxide and oxygen;
\item known C4 cycle metabolites alanine, aspartate, malate, PEP, and pyruvate;
\item the Calvin cycle intermediates glyceraldehyde 3-phosphate and 3-phosphoglycerate;
\item photorespiratory metabolites glycerate, glycolate, serine, and glycine;
\item nutrients sucrose, phosphate, nitrate, ammonia, sulfate and magnesium;
\item glutamate and 2-ketoglutarate; 
\item and cysteine and glutathione \cite{Burgener1998}. 
\end{itemize}
The inclusion of compounds involved in NAD-ME C4 or C3-C4 intermediate
photorespiratory carbon concentrating mechanism is not meant to
suggest such a system is necessarily active in maize but merely
reflects our knowledge that significant transport of those species
between mesophyll and bundle sheath can occur under at least some
circumstances.

\clearpage
\begin{suppfigure}
\includegraphics[width=\textwidth]{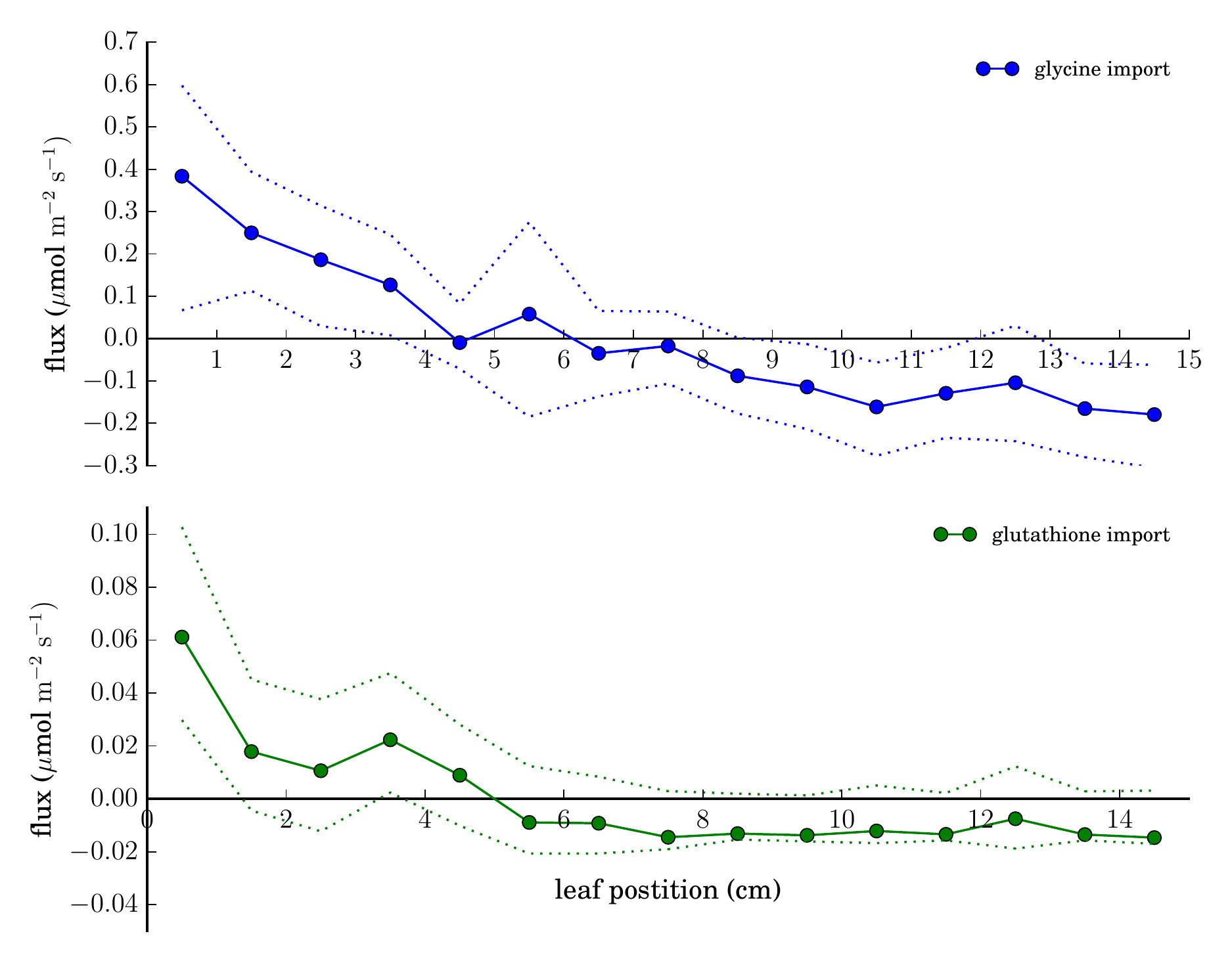}
\caption{
{\bf Phloem transport.} Transport of nitrogen (upper
  panel) and sulfur (lower panel) through the phloem in the
  best-fitting solution. Dotted lines indicate minimum and maximum
  predicted values consistent with an objective function value no more
  than 0.1\% worse than the optimum.
}
\label{S1_Figure} 
\end{suppfigure}

\begin{suppfigure}
\includegraphics[width=\textwidth]{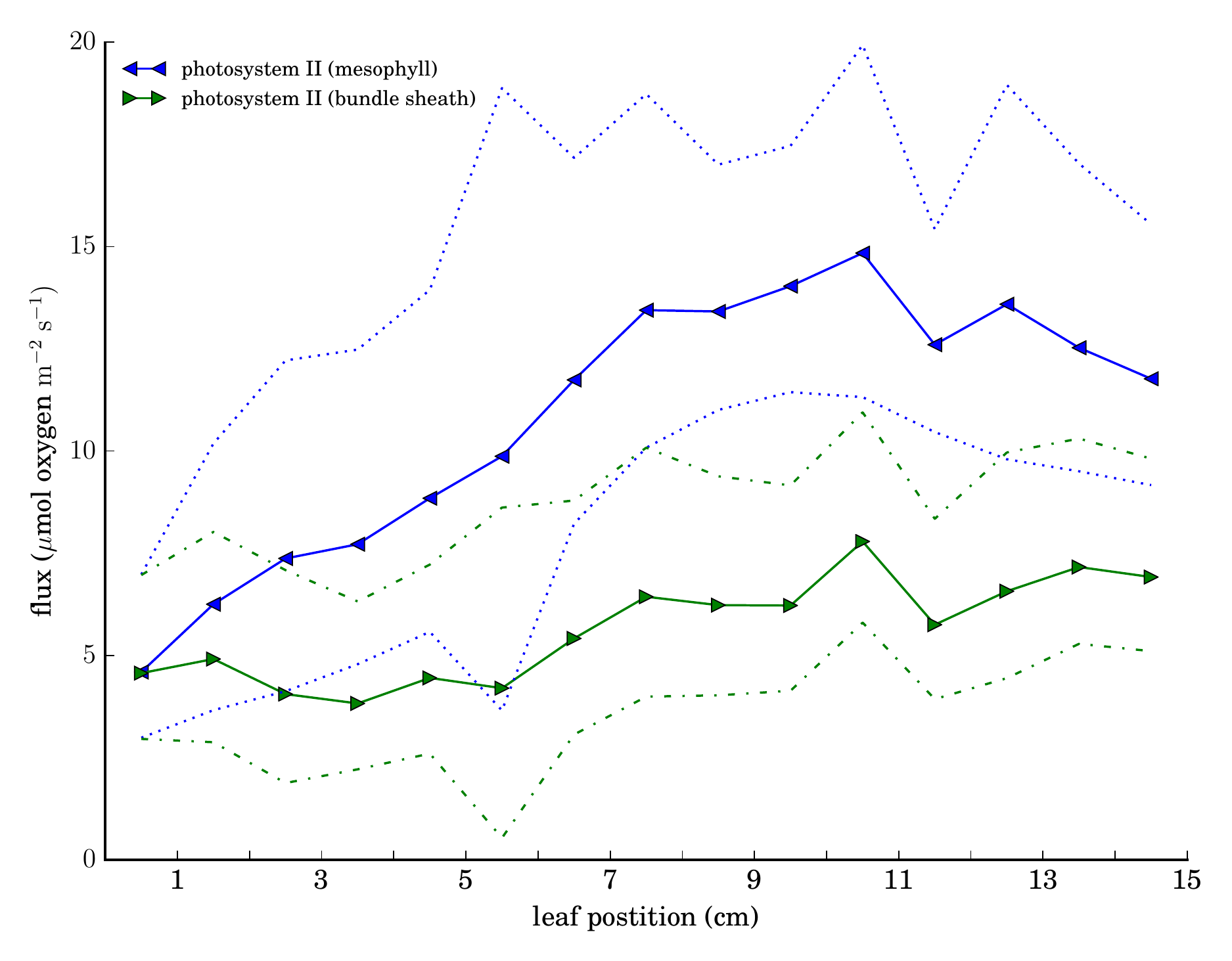}
\caption{
{\bf Photosystem II in mesophyll and bundle sheath.}
  Dashed and dotted lines indicate minimum and maximum predicted
  values consistent with an objective function value no more than
  0.1\% worse than the optimum.
}\label{S2_Figure}
\end{suppfigure}

\begin{suppfigure}
\includegraphics[width=\textwidth]{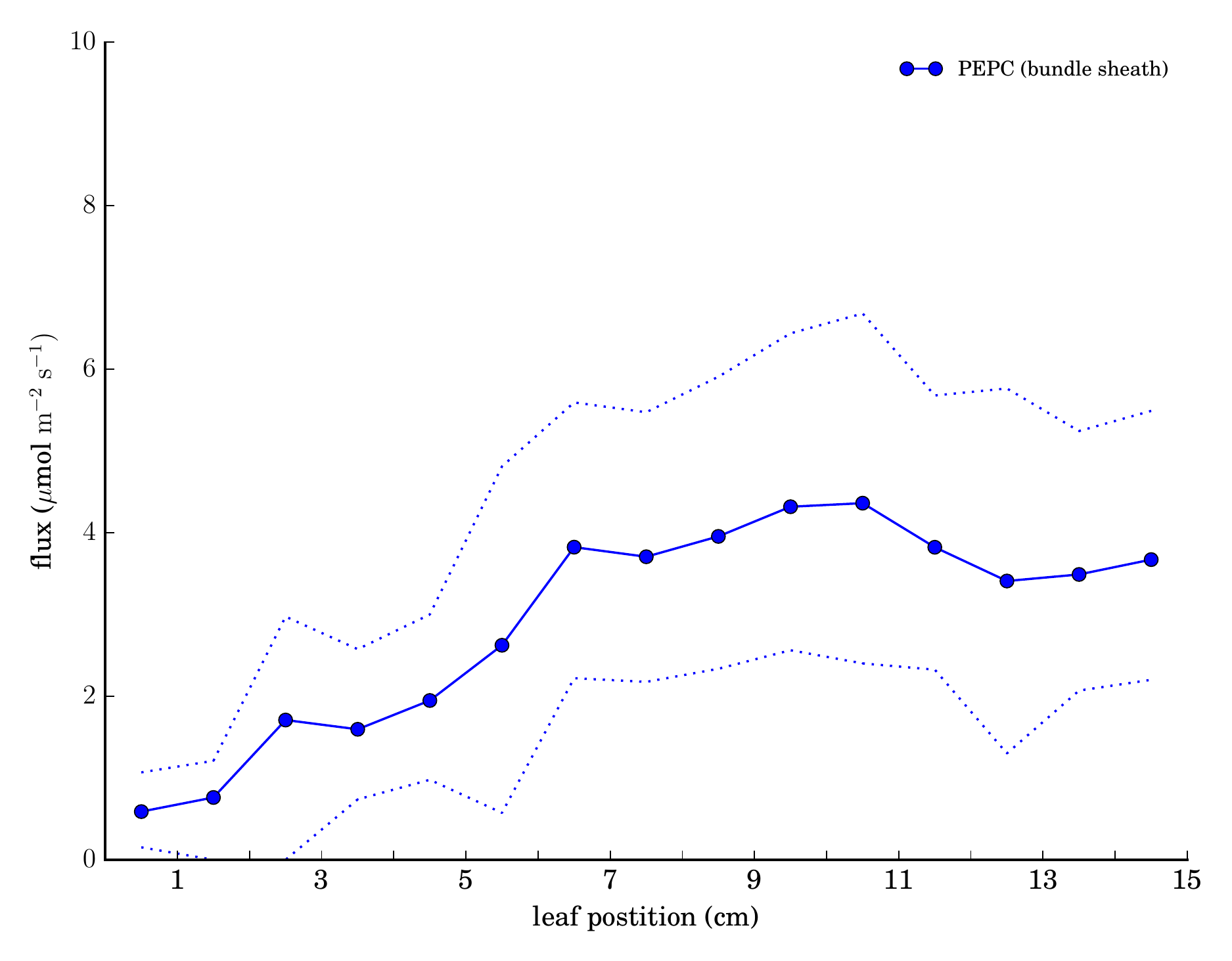}
\caption{{\bf Bundle sheath PEPC flux
  in the best-fitting solution.} Dotted lines indicate minimum and
  maximum predicted values consistent with an objective function value
  no more than 0.1\% worse than the optimum.}
\label{S3_Figure} 
\end{suppfigure}

\begin{suppfigure}
\includegraphics[width=\textwidth]{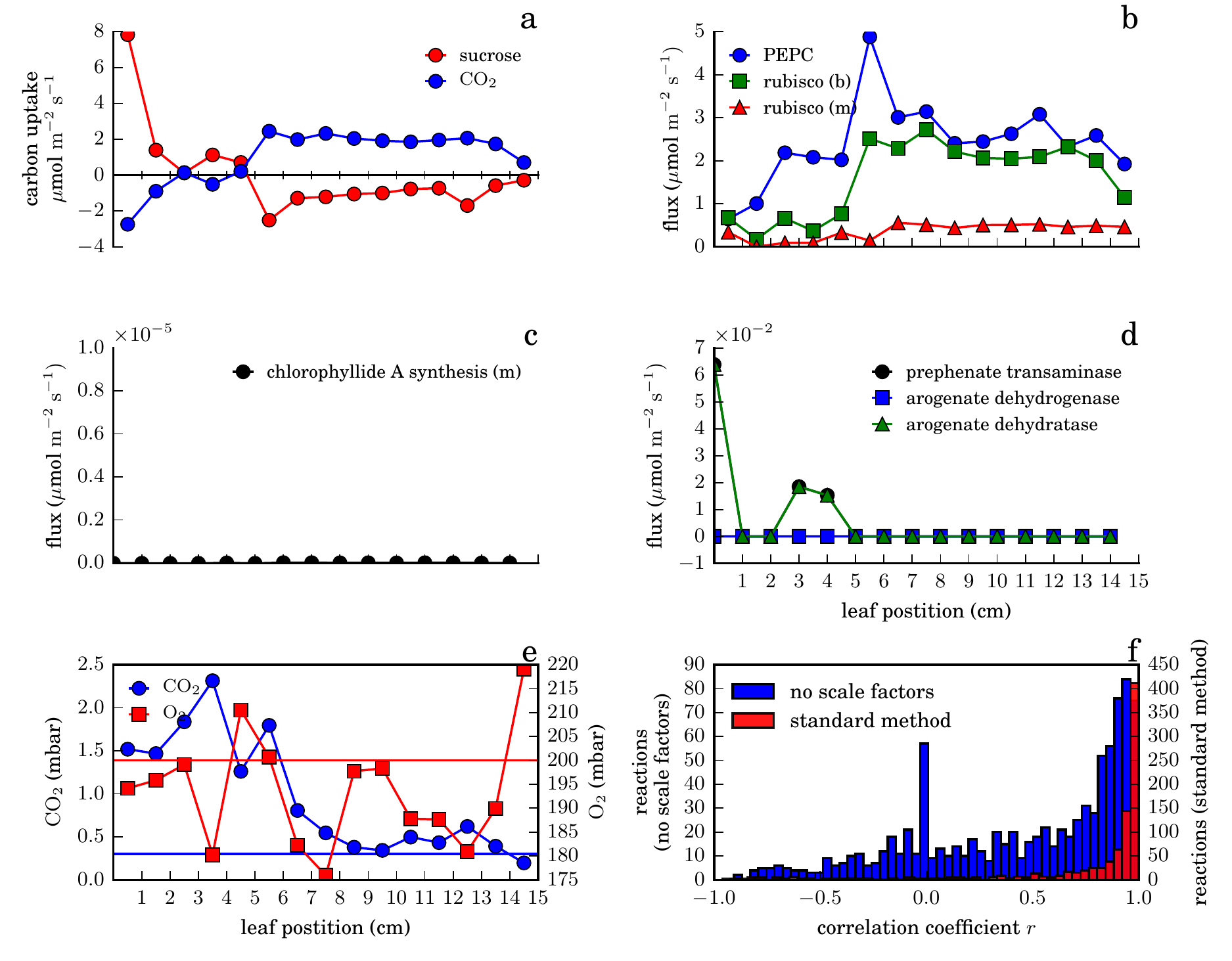}
\caption{
{\bf Summary of predictions for the gradient model
    using the least-squares method without per-reaction scale
    factors.} In eq. (\ref{fitting_no_abs}), $s_i=0$ for all reactions
  $i$.  (a) Sucrose and CO\textsubscript{2} uptake rates (compare to
  figure \ref{sourcesinkfig}a). (b) Rates of carboxylation by PEPC
  and Rubisco (compare to figure \ref{c4fig}b). (c) Predicted rate for
  the reactions of the chlorophyllide A synthesis pathway (compare to
  figure \ref{pathwayfig}b.) (d) Predicted rates at the arogenate
  branch point (compare to figure \ref{pathwayfig}d). (e) Predicted
  oxygen and carbon dioxide levels in the bundle sheath, with straight
  lines showing mesophyll levels (compare to figure \ref{c4fig}d). (f)
  Distribution of correlation coefficients between data and predicted
  fluxes for each reaction.  (blue, this method; red, standard
  method.)  Correlation coefficients for reactions with zero predicted
  flux are taken to be zero, resulting in the visible peak in the
  histogram.}\label{S4_Figure} 
\end{suppfigure}

\begin{suppfigure}
\includegraphics[width=\textwidth]{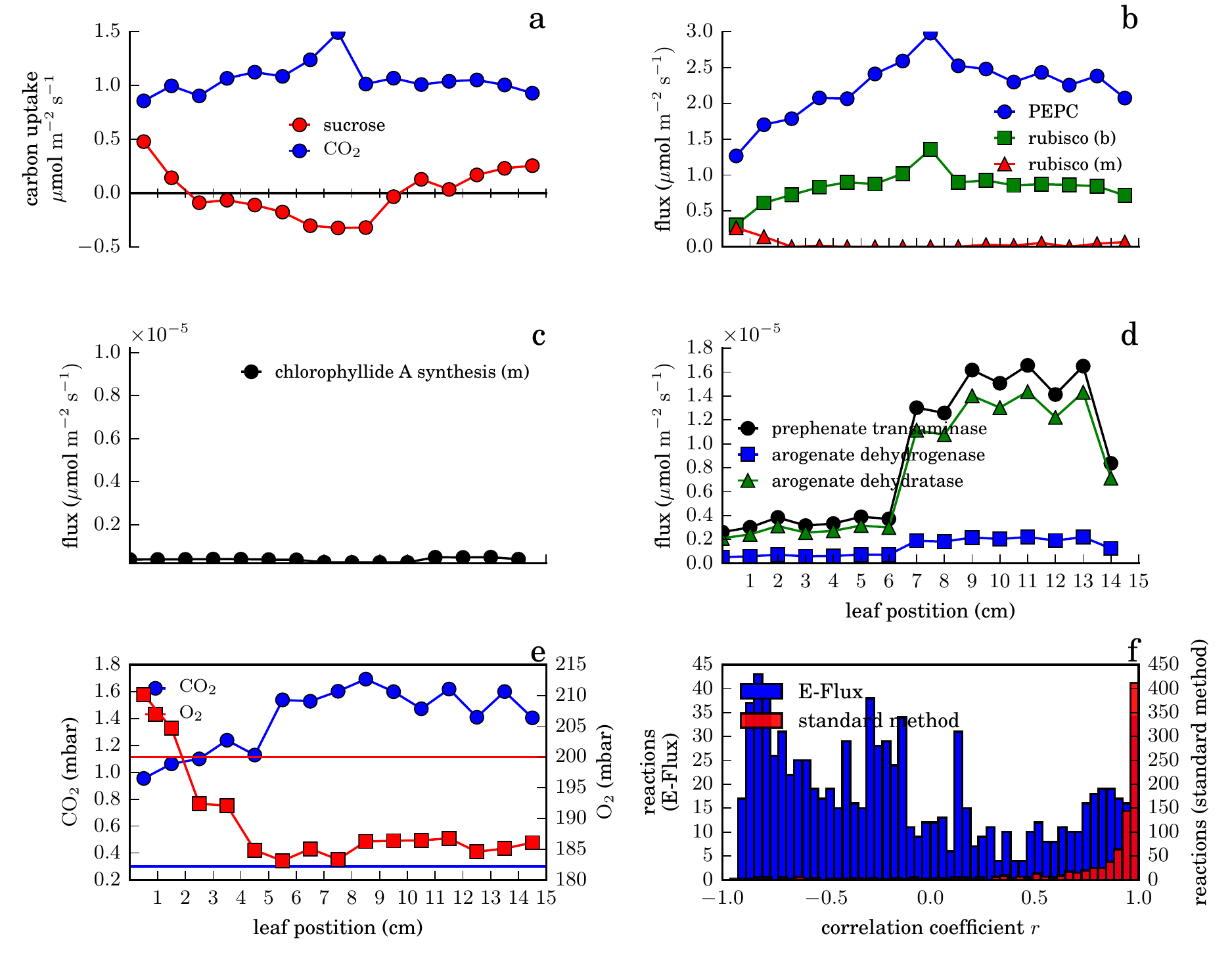}
\caption{
{\bf Summary of predictions for the gradient model
    using the E-Flux method.} For explanation of each panel, see \autoref{S4_Figure}. 
} \label{S5_Figure} 
\end{suppfigure}

\begin{suppfigure}
\includegraphics[width=\textwidth]{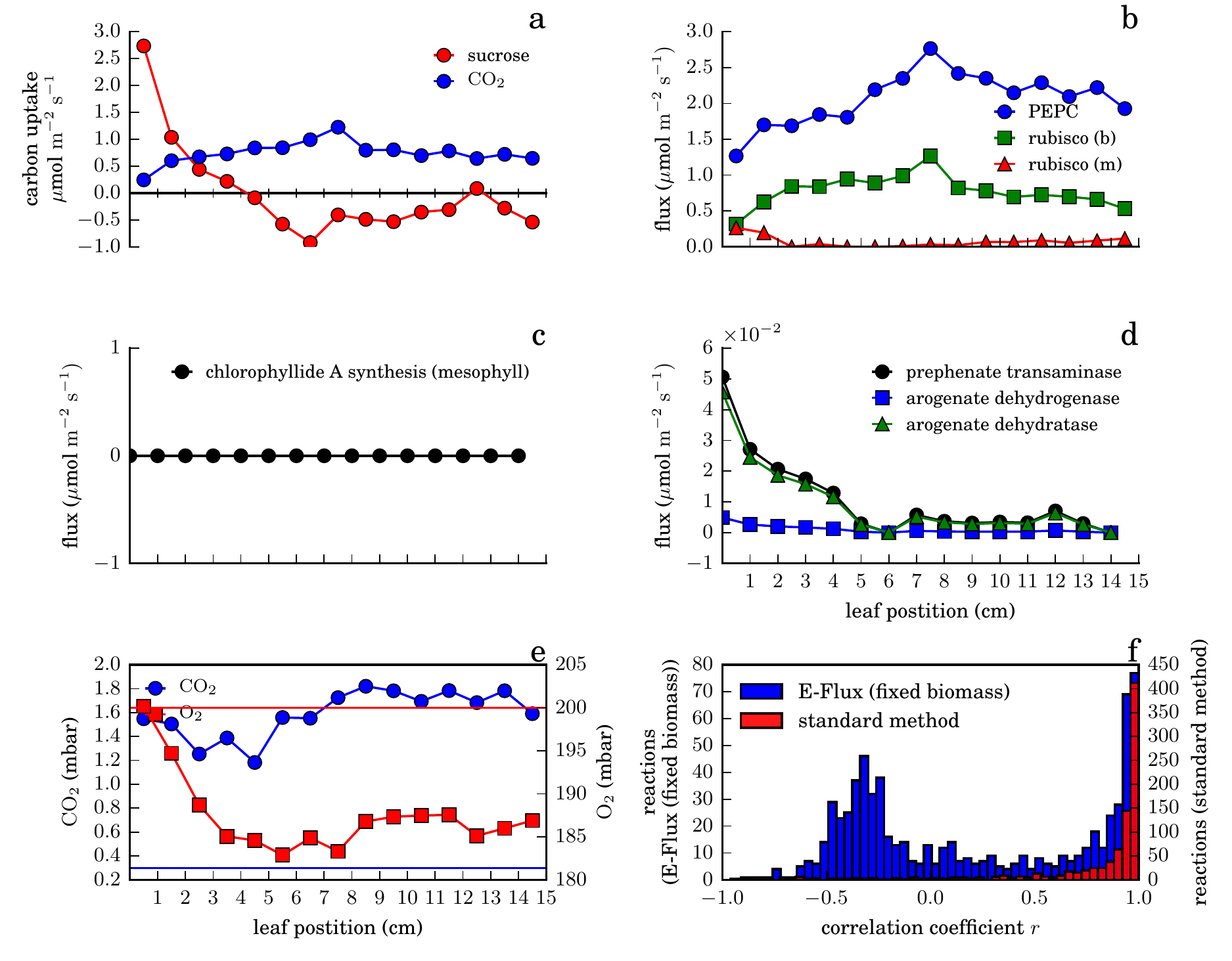}
\caption{ {\bf Summary of
    predictions for the gradient model using the E-Flux method with
    fixed biomass composition.} The biomass composition is fixed to
  that used by iRS1563, as adapted (see \nameref{S9_Appendix}).  For
  explanation of each panel, see \autoref{S4_Figure}. Note that the
  chlorophyllide A synthesis pathway is blocked when the fixed biomass
  composition is used. } \label{S6_Figure}
\end{suppfigure}

\begin{suppfigure}
\includegraphics[width=\textwidth]{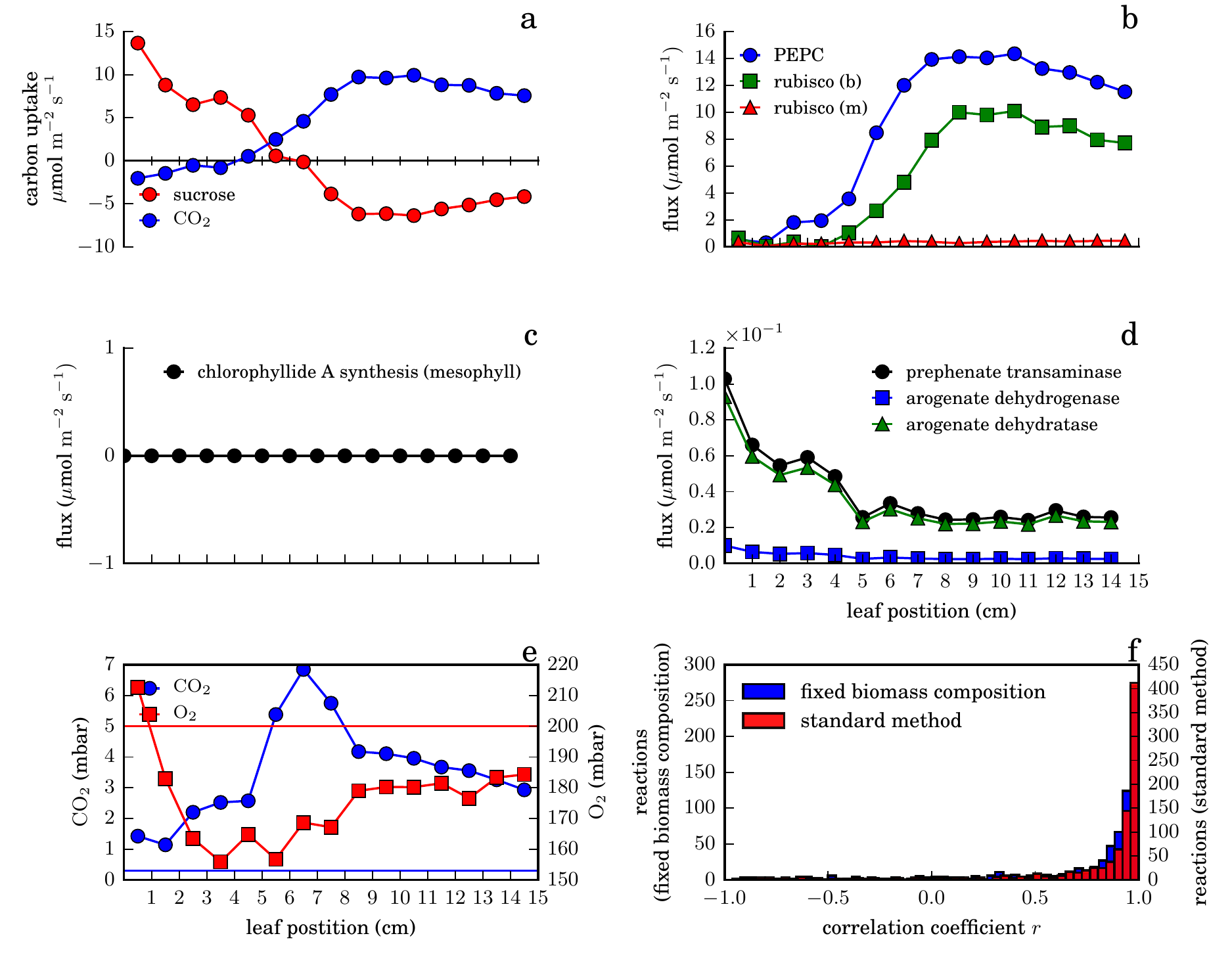}
\caption{{\bf Summary of predictions for
  the gradient model with fixed biomass composition.}  For explanation
  of each panel, see \autoref{S4_Figure}.  Note that the
  chlorophyllide A synthesis pathway is blocked when the fixed biomass
  composition is used.}\label{S7_Figure}
\end{suppfigure}

\begin{suppfigure}
\includegraphics[width=\textwidth]{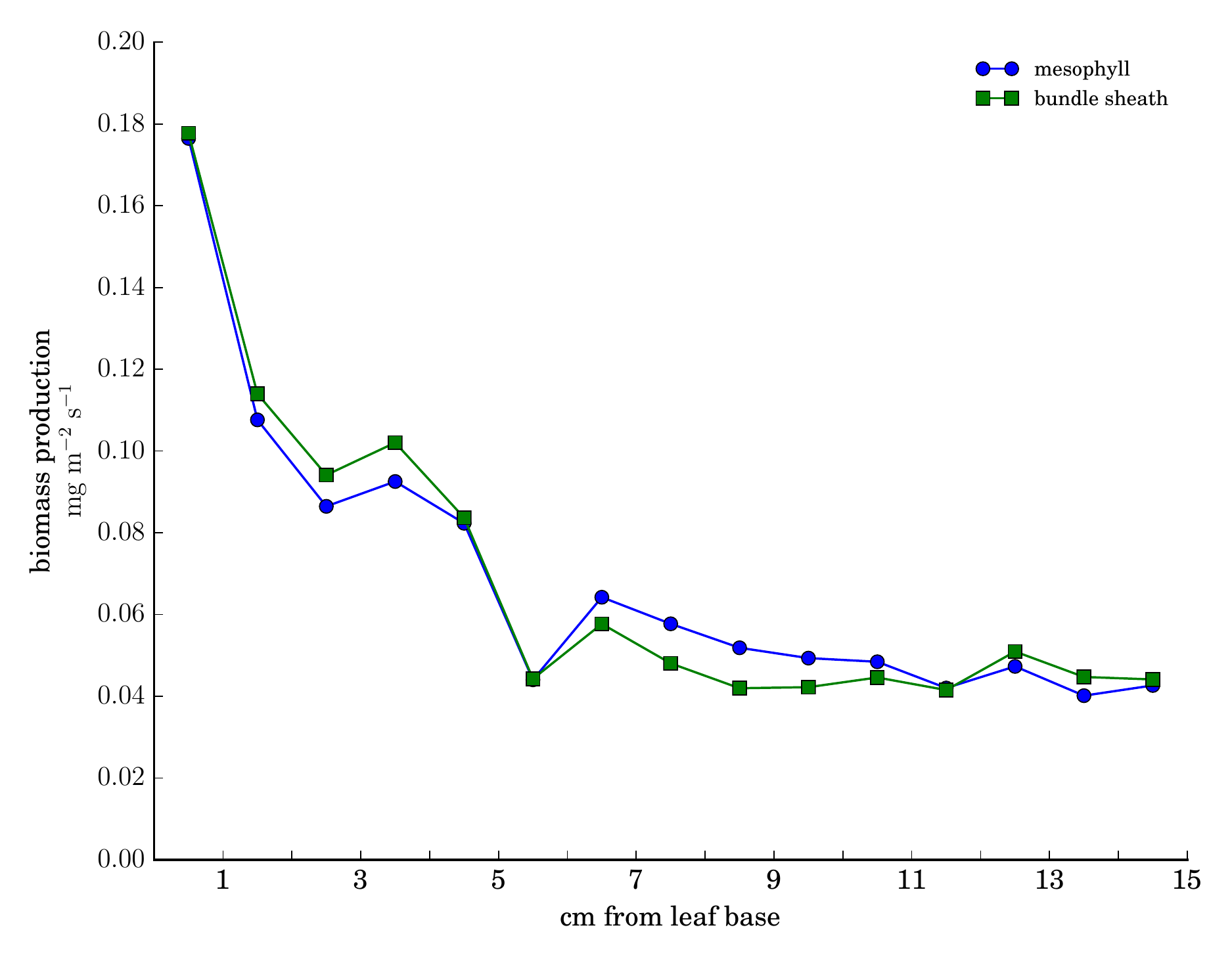}
\caption{ {\bf Predicted biomass
  production rates in mesophyll and bundle sheath cells with fixed
  biomass composition.}}\label{S8_Figure}
\end{suppfigure}

\begin{supptable}
\begin{adjustbox}{max width=1.25\textwidth,center}
{\small
\begin{tabular}{l|l|l}
reaction & name in model & associated genes \\
\hline
malate dehydrogenase (NADP)  & \texttt{MALATE\_DEHYDROGENASE\_NADP\_\_RXN\_chloroplast} & 1 \\
alanine aminotransferase & \texttt{ALANINE\_AMINOTRANSFERASE\_RXN} & 10 \\
aspartate aminotransferase & \texttt{ASPAMINOTRANS\_RXN} & 7 \\
NAD-malic enzyme & \texttt{EC\_1\_1\_1\_39} & 2 \\
NADP-malic enzyme (cytosol) &\texttt{MALIC\_NADP\_RXN} & 4 \\
NADP-malic enzyme (chloroplast) & \texttt{MALIC\_NADP\_RXN\_chloroplast} & 2 \\
PEPCK & \texttt{PEPCARBOXYKIN\_RXN} & 6 \\
PPDK & \texttt{PYRUVATEORTHOPHOSPHATE\_DIKINASE\_RXN\_chloroplast} & 2 \\
adenylate kinase & \texttt{ADENYL\_KIN\_RXN\_chloroplast} & 6 \\
pyrophosphatase & \texttt{INORGPYROPHOSPHAT\_RXN\_chloroplast} & 2 \\
\end{tabular}
}
\end{adjustbox}
\vskip 0.25cm
  \caption{{\bf Detailed parameters
    contributing to the effective PEP regeneration rate: reactions in
    the genome-scale model which contribute to the effective maximum
    PEP regeneration capacity, and the number of genes associated with
    each.} In addition to the reactions listed, transport capacities
  of pyruvate, PEP, alanine, aspartate and malate across the
  plasmodesmata and pyruvate, PEP, malate and oxaloacetate across the
  chloroplast inner membrane could limit this rate; the model
  currently associates no genes with these transport reactions.}\label{S10_Table}
\vskip 10cm
\end{supptable}

\begin{suppfigure}
\includegraphics[width=\textwidth]{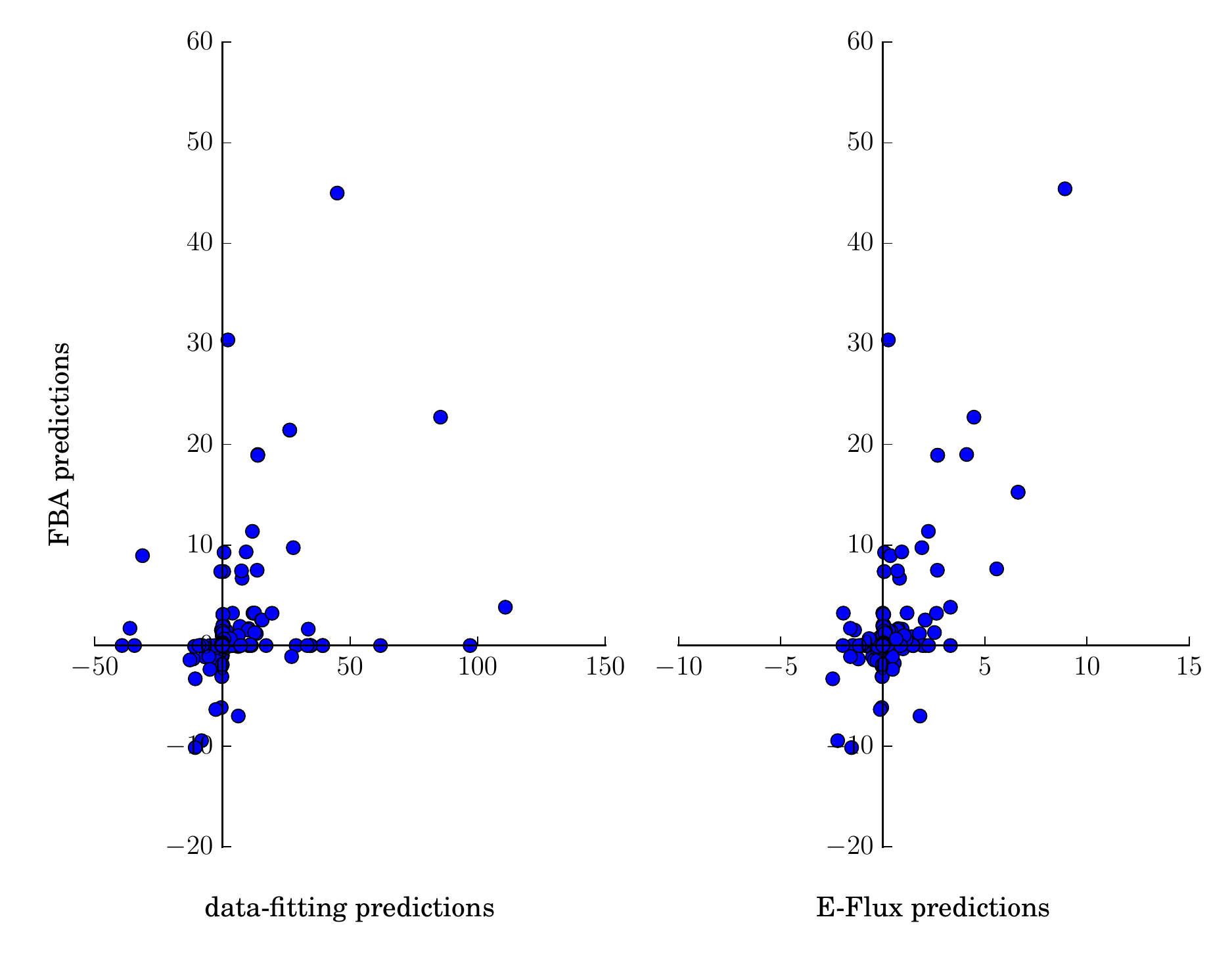}
\caption{ {\bf Predicted variable values in an FBA calculation that
    does not incorporate expression data, compared to the best-fit and
    E-Flux methods.}  The FBA calculation minimizes total flux while
  achieving the same total rate of CO\textsubscript{2} assimilation as
  predicted at the tip of the leaf in the fitting results. Left panel,
  FBA reaction rates vs. reaction rates predicted at the tip of the
  leaf in the best-fitting solution; right panel, FBA reaction rates
  vs. reaction rates predicted at the tip of the leaf by the E-Flux
  method. Axis limits exclude a small number of reactions of
  particularly large flux.  Fluxes in \fluxunit.}\label{S18_Figure}
\end{suppfigure}
\clearpage

\section*{Index to additional supporting information files}
Except as noted, these are available as arXiv ancillary files.
  \subsection*{S11 Model} \label{S11_Model} {\bf iEB5204 in SBML format.}
  \subsection*{S12 Model} \label{S12_Model} {\bf iEB2140 in SBML format.}
  \subsection*{S13 Model} \label{S13_Model} {\bf iEB2140x2 in SBML format.}
  \subsection*{S14 Protocol} \label{S14_Protocol} {\bf Source code for the nonlinear constraint-based modeling package fluxtools.} Available at 
\url{http://github.com/ebogart/fluxtools}.
  \subsection*{S15 Protocol} \label{S15_Protocol} {\bf Source code and input files for the calculations discussed above.}  Available at \url{http://github.com/ebogart/multiscale_c4_source}.
  \subsection*{S16 Table} \label{S16_Table} {\bf Predicted variable values along the leaf gradient.}
  \subsection*{S17 Table} \label{S17_Table} {\bf Upper and lower bounds on predicted values of selected variables along the leaf gradient, from FVA calculations.}

\end{document}